\documentclass[utf8]{frontiersSCNS} % for Science, Engineering and Humanities and Social Sciences articles

\setcitestyle{square} % for Physics and Applied Mathematics and Statistics articles
\usepackage{url,hyperref,lineno,microtype,subcaption}
\usepackage[onehalfspacing]{setspace}

\usepackage{titlesec, blindtext, color}
	\definecolor{light-blue}{rgb}{0.8,0.85,1}
	\definecolor{airforceblue}{rgb}{0.36, 0.54, 0.66}
	\definecolor{azure}{rgb}{0.0, 0.5, 1.0}
	\definecolor{bleudefrance}{rgb}{0.19, 0.55, 0.91}
	\definecolor{blue(munsell)}{rgb}{0.0, 0.5, 0.69}
	\definecolor{darkmidnightblue}{rgb}{0.0, 0.2, 0.4}
	\definecolor{steelblue}{rgb}{0.27, 0.51, 0.71}
	\definecolor{tealblue}{rgb}{0.21, 0.46, 0.53}
	\definecolor{yaleblue}{rgb}{0.06, 0.3, 0.57}
	\definecolor{applered}{rgb}{0.89, 0.02, 0.17}

\newcommand{\radyn}{\texttt{RADYN}}
\newcommand{\fp}{\texttt{FP}}
\newcommand{\radynfp}{\texttt{RADYN+FP}}
\newcommand{\radynarcade}{\texttt{RADYN\_Arcade}}
\newcommand{\msradyn}{\texttt{MS\_RADYN}}

\newcommand{\preft}{\texttt{PREFT}}
\newcommand{\hydrad}{\texttt{HYDRAD}}
\newcommand{\flarix}{{\texttt{FLARIX}}}

\newcommand{\rh}{\texttt{RH}}
\newcommand{\rhpar}{\texttt{RH15D}}
\newcommand{\mali}{\texttt{MALI}}

\newcommand\ion[2]{#1$\;${%
\ifx\@currsize\normalsize\small \else
\ifx\@currsize\small\footnotesize \else
\ifx\@currsize\footnotesize\scriptsize \else
\ifx\@currsize\scriptsize\tiny \else
\ifx\@currsize\large\normalsize \else
\ifx\@currsize\Large\large
\fi\fi\fi\fi\fi\fi
\rmfamily\textsc{#2}}\relax}% 

\newcommand{\gskfont}{
  \bfseries
  \color{applered}
}
\DeclareTextFontCommand{\gsk}{\gskfont}

\def\keyFont{\fontsize{8}{11}\helveticabold }
\def\firstAuthorLast{Kerr, G.S.} 
\def\Authors{Graham S. Kerr\,$^{1,2,*}$}
\def\Address{$^{1}$Department of Physics, The Catholic Univeristy of America, 620 Michigan Avenue, Northeast, Washington D.C. 20064, USA \\
$^{2}$NASA Goddard Space Flight Center, Heliophysics Sciences Division, Code 671, 8800 Greenbelt Road, Greenbelt MD 20771, USA }

\def\corrAuthor{Graham S. Kerr}

\def\corrEmail{graham.s.kerr@nasa.gov, kerrg@cua.edu}

\begin{document}
\onecolumn
\firstpage{1}

\title[Solar Flare Loop Models \& IRIS, Paper 2]{Interrogating Solar Flare Loop Models with IRIS Observations 2: Plasma Properties, Energy Transport, and Future Directions} 

\author[\firstAuthorLast ]{\Authors} %This field will be automatically populated
\address{} %This field will be automatically populated
\correspondance{} %This field will be automatically populated

\extraAuth{}% If there are more than 1 corresponding author, comment this line and uncomment the next one.
%\extraAuth{}

\maketitle

%%%%%%%%%%%%%%%%%%%%%%%%%%%%%%%%%%%%%%%%%%%%%
%% ABSTRACT
%%%%%%%%%%%%%%%%%%%%%%%%%%%%%%%%%%%%%%%%%%%%%

\begin{abstract}

\section{}
During solar flares a tremendous amount of magnetic energy is released and transported through the Sun's atmosphere and out into the heliosphere. Despite over a century of study, many unresolved questions surrounding solar flares are still present. Among those are how does the solar plasma respond to flare energy deposition, and what are the important physical processes that transport that energy from the release site in the corona through the transition region and chromosphere? Attacking these questions requires the concert of advanced numerical simulations and high spatial-, temporal-, and spectral- resolution observations. While flares are 3D phenomenon, simulating the NLTE flaring chromosphere in 3D and performing parameter studies of 3D models is largely outwith our current computational capabilities. We instead rely on state-of-the-art 1D field-aligned simulations to study the physical processes that govern flares. Over the last decade, data from the Interface Region Imaging Spectrograph (IRIS) have provided the crucial observations with which we can critically interrogate the predictions of those flare loop models. Here in Paper 2 of a two-part review of IRIS and flare loop models, I discuss how forward modelling flares can help us understand the observations from IRIS, and how IRIS can reveal where our models do well and where we are likely missing important processes, focussing in particular on the plasma properties, energy transport mechanisms, and future directions of flare modelling.

\tiny
 \keyFont{ \section{Keywords:} solar flares, solar atmosphere, solar chromosphere, UV radiation, Numerical methods, Radiation Transfer} %All article types: you may provide up to 8 keywords; at least 5 are mandatory.
\end{abstract}

%%%%%%%%%%%%%%%%%%%%%%%%%%%%%%%%%%%%%%%%%%%%%
%% INTRODUCTION
%%%%%%%%%%%%%%%%%%%%%%%%%%%%%%%%%%%%%%%%%%%%%

\section{Introduction}

\subsection{Solar Flares}
Solar flares are transient, broadband brightenings to the Sun's radiative output following the liberation of a tremendous amount of energy (up to $10^{32}$~erg, or larger: \citealt{2012ApJ...759...71E,2015ApJ...802...53A}) during magnetic reconnection \citep[e.g.][]{2002A&ARv..10..313P, 2011LRSP....8....6S,2013A&A...555A..77J,2012ApJ...759...71E}. It is thought that this energy is subsequently transported predominately in the form of non-thermal particles. We primarily consider non-thermal electrons\footnote{Though ions likely carry significant amounts of energy \citep[e.g.][]{2012ApJ...759...71E}, the flare community primarily considers electrons since their signatures are easier to infer from hard X-rays. Flare accelerated protons and heavier ions are harder to constrain with current observational capabilities}, accelerated during the reconnection process. Once they reach the denser lower solar atmosphere they thermalise via Coulomb collisions \citep[e.g.][]{1971SoPh...18..489B}, heating and ionising the plasma and generating mass flows: chromospheric evaporation (upflowing material) and chromospheric condensations (downflowing material). Alternative mechanisms of energy transport in flares include non-thermal protons or heavier ions, thermal conduction following direct heating of the corona, and Alfv\'enic waves, discussed in more detail in Section~\ref{sec:mechs}.

There is unambiguous evidence for the presence of non-thermal particles in flares, due to the hard X-rays they produce via bremsstrahlung. Their ubiquitousness and the close spatial and temporal association with other flare emission (e.g. optical and UV) has bolstered the `electron-beam' model of solar flares. Observations of hard X-rays, e.g. from the Reuven Ramaty High Energy Solar Spectroscopic Imager \citep[RHESSI;][]{2002SoPh..210....3L}, can be used to infer the underlying non-thermal electron energy distribution, that itself can drive models of solar flares. There is a substantial body of literature describing the various characteristics of flares, and the means in which we observe them. I direct the reader towards the following reviews of flare observations, and flare particle acceleration and thermalisation: \cite{2008LRSP....5....1B,2011SSRv..159...19F,2011SSRv..159..107H,2011SSRv..159..301K,2011SSRv..159..357Z,2015SoPh..290.3399M}. The bulk of the flare radiative output originates from in chromosphere and transition region, making those regions important areas of study for their diagnostic potential regarding the plasma response to energy injection, and the energy transport and release process themselves. However, speaking candidly, this potential has been somewhat squandered by the lack of routine high spatial-, temporal-, and  spectral- observations of the chromosphere and transition region at UV wavelengths during flares (crucially, we lacked routine imaging spectroscopy of the flare chromosphere). That observational gap has fortunately been plugged following the launch of the Interface Region Imaging Spectrograph \citep[IRIS;][]{2014SoPh..289.2733D} in 2013, that now gives us an unprecedented view of the flaring chromosphere and transition region, yielding crucial new insights. Given the complex environment of these particular layers, parallel efforts to forward model the flaring lower atmosphere, and its impacts on the flaring corona, are required to make substantial progress in understanding the physics at play in flares.

This is the second paper in a two part review of how solar flare loop models in concert with IRIS observations have improved our understanding of solar flares. Between both parts I hope to emphasise that it is only by attacking the problem of flare physics via the combination of high quality observations and state-of-the-art models, that include the pertinent physical processes, that we can make rapid progress. Overall I aim to show: (1) how modelling has helped interpret the IRIS observations; (2) how IRIS observations have been used to interrogate and validate model predictions; and (3) how, when models fail to stand up to the stubborn reality of those observations, IRIS has led to model improvements.  In Paper 1 of this review \citep{kerr2022_irisflarereview_p1} I provided a detailed overview of each numerical code, and discussed what we have learned from the study of Doppler motions from IRIS in the context of the non-thermal electron beam driven flare model. Also in Paper 1 is a more extensive introduction to solar flares. Here in Paper 2 I demonstrate how we have used the combination of IRIS and flare loop modelling to learn about plasma properties and flare energy transport mechanisms, and provide some thoughts on  future directions. 

IRIS is a NASA Small Explorer mission that has observed many hundreds of flares, including dozens of M and X class events. Both images (via the slit-jaw imager, SJI) and spectra (via the slit-scanning spectrograph, SG) are provided in the far- and near-UV (FUV \& NUV), with a spatial resolution $0.33-0.4$~arcseconds. High cadences are achievable, as low as $1$~s but more generally a few seconds to tens of seconds. The strongest lines observed are \ion{Mg}{ii} h 2803~\AA\ \& k 2796~\AA\ lines (chromosphere), \ion{C}{ii} 1334~\AA\ \& 1335~\AA\ and \ion{Si}{iv} 1394~\AA\ \& 1403~\AA\ (transition region), and \ion{Fe}{xxi} 1354.1~\AA\ ($\sim11$~MK plasma), with numerous other lines observed in the three passbands $[1332-1358]$~\AA, $[1389-1407]$~\AA, and $[2783-2834]$~\AA. The spectral resolution is $\sim53$~m\AA\ in the NUV and $\sim26$~m\AA\ in the FUV. The SJI observes at $2796\pm4$~\AA\ (\ion{Mg}{ii} k), $2832\pm4$~\AA\ (\ion{Mg}{ii} wing plus quasi-continuum), $1330\pm55$~\AA\ (\ion{C}{ii}), and $1400\pm55$~\AA\ (\ion{Si}{iv}). See \cite{2021SoPh..296...84D} for a review of the various successes over the first near-decade of IRIS observations. 

The models employed to study flares are generally field-aligned (1D) numerical codes \citep[though there are some exceptions, e.g. the 3D radiative magnetohydrodynamic, MHD, model of ][]{2019NatAs...3..160C}. These codes are nimble enough to be run on timescales that make performing parameter studies of flare energy transport processes a tractable activity, and they allow us to include the relevant physical processes at the required spatial resolution (down to sub-metres) that is not yet feasible in 3D RMHD simulations. I focus on the \radyn, \hydrad, \flarix, and \preft\ models here. A brief overview is presented below but see Paper 1 for a full description of each code. 

\subsection{Summary of the Models}
The hydrodynamic field-aligned codes \hydrad\ \citep{2003A&A...401..699B,2003A&A...407.1127B,2013ApJ...778...76R,2019ApJ...871...18R}, \radyn\ \citep{1992ApJ...397L..59C,2002ApJ...572..626C,1997ApJ...481..500C,1999ApJ...521..906A,2005ApJ...630..573A,2015ApJ...809..104A}, and \flarix\ \citep{2009A&A...499..923K,2010ITPS...38.2249V,2016IAUS..320..233H} are now well established and widely used in the solar flare community. These codes solve the equations describing the conservation of mass, momentum, charge, and energy in a single field-aligned magnetic strand rooted in the photosphere and stretching out to include the chromosphere, transition region, and corona. \hydrad\ and \radyn\ use an adaptive grid where the size of the grid cells can vary to allow shocks and steep gradients in the atmosphere to be resolved as required (with \hydrad\ varying the the number of grid cells also), while \flarix\ uses a fixed, but optimised, grid with $\sim2000$ points. The codes have various similarities and differences as regards treatment of radiation and flare energy transport, and with the numerical approaches themselves.

 All three simulate the response of the atmosphere to injection of energy, typically via a beam of non-thermal electrons (but flare-accelerated ions can be included too). \radyn\ uses a Fokker-Plank treatment to model the evolution of the non-thermal electron distribution as a function of time (including return current effects), that was recently updated to employ the standalone state-of-the-art non-thermal particle transport code \fp \footnote{\url{https://github.com/solarFP/FP}} \citep[][]{2020ApJ...902...16A}. \hydrad\ uses the analytic treatment of \cite{1978ApJ...224..241E} and \cite{1994ApJ...426..387H}, and \flarix\ uses a test-particle module that provides the time-dependent beam propagation including scattering terms. Dissipation of Alfv\'enic waves has also been recently implemented in both \hydrad\ and \radyn\ \citep{2016ApJ...818L..20R,2018ApJ...853..101R,2016ApJ...827..101K}, and all codes can include \textsl{ad-hoc} time dependent heating.

Each code has been conceived and developed to focus on particular details of the flaring plasma physics problem. \radyn\ and \flarix\ are radiation hydrodynamic codes which couple the hydrodynamic equations to the non-LTE (NLTE) 1D radiative transfer and time-dependent non-equilibrium atomic level population equations, for elements important for chromospheric energy balance. 
\radyn\ considers H, He \& Ca, with Mg also sometimes included, whereas \flarix\ considers H, Ca, and Mg (with plans to update the code to include He). Continua from other species are treated in LTE as background metal opacities. Optically thin losses are included by summing all transitions from the \texttt{CHIANTI} atomic database \citep{1997A&AS..125..149D,2015A&A...582A..56D,2021ApJ...909...38D}\footnote{\radyn\ and \flarix\ currently uses CHIANTI V8, and \hydrad\ V10, but the version should be noted in the relevant studies.} apart from those transitions solved in detail. Additional backwarming and photoionisations by soft X-ray, extreme ultraviolet, and ultraviolet radiation is included. Both currently employ the assumption of complete frequency redistribution (CRD)\footnote{As stated in Paper 1: CRD assumes that the wavelength of a scattered/emitted photon is uncorrelated to the wavelength at which it was absorbed, due to collisions (e.g. photons absorbed in the line wings may be redistributed and emitted at a wavelength in the line core). However, in relatively low-density environments such as the chromosphere there may be an insufficient number of elastic collisions such that the scattered photon has a wavelength that is correlated to that of the absorbed photon. Photons absorbed in the line wings are re-emitted in the wings, where it easier to escape. This is the partial frequency redistribution (PRD) scenario. CRD has a frequency independent source function, whereas PRD has a frequency dependent source function and the absorption profile does not equal the emission profile. See discussions in \cite{1982JQSRT..27..593H}, and \cite{2001ApJ...557..389U,2002ApJ...565.1312U}.} when solving the radiation transport problem, so that post-processing via other radiation transport codes such as \rh, \rhpar, or \mali\ is required. In \radyn\ and \flarix\ the loop is modelled as one leg of a symmetric flux tube. \radyn\ also allows to calculate \textsl{aposteriori} (i.e. with no feedback on the plasma equations of mass, momentum, and energy) the time-dependent non-equilibrium populations and radiation transport of a desired ion via the minority species version of that code, \msradyn\ \citep[][]{2003ApJ...597.1158J,2019ApJ...885..119K,2019ApJ...871...23K}.  

\hydrad\ does not solve the detailed optically thick radiation transport and atomic level population equations, instead employing  approximations of chromospheric radiation losses. Losses from H, Ca and Mg are included via the approach of \cite{2012A&A...539A..39C}. The code has also recently adopted a more accurate method for computing NLTE H populations following the prescription of \cite{1999MsT..........1S} which approximates the radiation field in the chromosphere \citep{2019ApJ...871...18R}. Ion population equations, however, are solved self-consistently in full non-equilibrium ionisation (NEI) for any desired element, returning a more accurate calculation of the optically thin radiative losses and spectral synthesis of optically thin lines using those ion fractions. While the treatment of optically thick radiation is less robust than in \radyn\ or \flarix, \hydrad\ has the advantage of being significantly less computationally demanding. Other important differences are that \hydrad\ features a multi-fluid plasma that treats the electron and hydrogen temperatures separately, it solves a full length flux tube (foot-point to foot-point) of arbitrary geometry (e.g. based on a magnetic field extrapolation) and includes effects due to cross-sectional area expansion (varying inversely with the magnetic field strength), which has been shown to play an important role in dynamics \citep[][]{2022ApJ...927..103R}.

\preft\ \citep{2010ApJ...718.1476G,2011ApJ...740...73L,2015ApJ...813..131L,2016ApJ...833..211L} is a rather different code than the other three, and is a powerful tool to study the impact of magnetic loop dynamics during flares. It is a 1D MHD code that solves the thin flux tube (TFT) equations. The tube is initialised at the instant after a localised reconnection process within the current sheet has linked sections of equilibrium tubes from opposite sides of the current sheet.  No further reconnection occurs, and any heating from the initialising event is neglected.    In its subsequent evolution, the tube retracts under magnetic tension releasing magnetic energy and converting it to bulk kinetic energy in flows which include a component parallel to the tube. The collision between the parallel components generates a pair of propagating slow magnetosonic shocks, which resemble gas dynamic shocks as they must in the parallel limit. Radiative losses are optically thin, and normally an isothermal, but gravitationally stratified, chromosphere is included mostly as a mass reservoir. Solutions of the TFT equations show that thermal conduction carries heat away from the shocks, drastically altering the temperature and density of the post-flare plasma \citep[][]{2011ApJ...740...73L,2015ApJ...813..131L,2016ApJ...833..211L,2020ApJ...894..148U}.
 
 %%%%%%%%%%%%%%%%%%%%%%%%%%%%%%%%%%%%%%%%%%%%%
%% PLASMA PROPERTIES
%%%%%%%%%%%%%%%%%%%%%%%%%%%%%%%%%%%%%%%%%%%%%

\section{Plasma Properties in the Flaring Atmosphere}\label{sec:plasmaprops}
Since much of emission from the flaring chromosphere is optically thick, extracting meaningful information about the plasma properties is difficult and often requires forward modelling from flare simulations in order to interpret observations. Even in the corona where emission is optically thin, modelling is required. In this section I present some examples of where flare models have shed light on conditions in the flare atmosphere.

\subsection{Understanding the Flaring Chromosphere via \ion{Mg}{ii}}

\begin{figure}[h]
\begin{center}
\includegraphics[width=\textwidth]{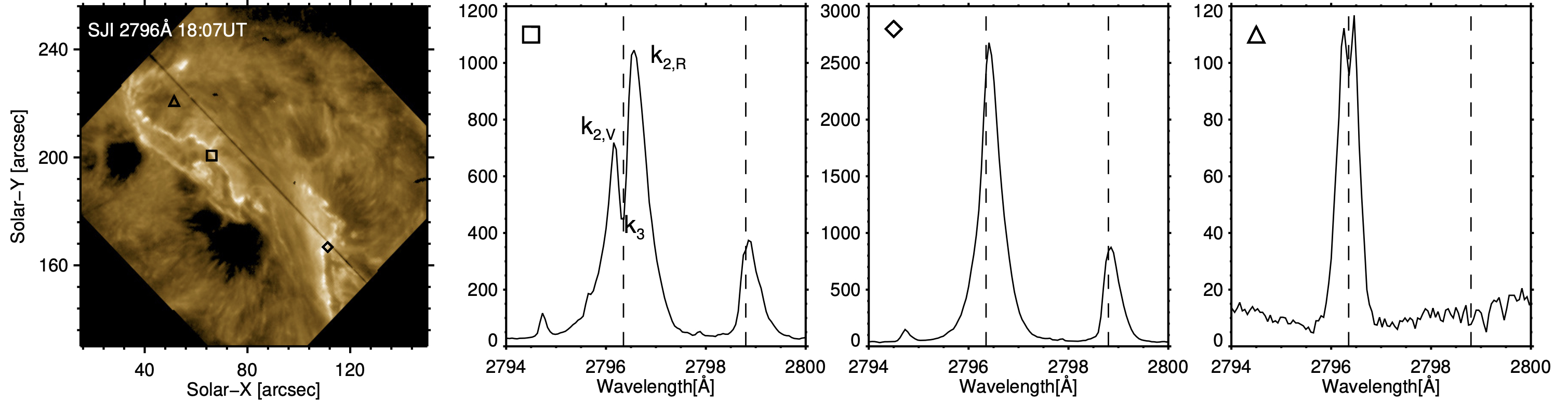}% This is a *.eps file
\end{center}
\caption{An illustration of the \ion{Mg}{ii} profiles as observed by IRIS. The map on the left shows a flare image from the SJI. The spectra in the other three panels come from the locations identified by symbols on the map. The square (second panel) is a profile from the leading edge of a flare ribbon, where the different line components are labelled. The diamond (third panel) is a source in the middle of a flare ribbon. The triangle (fourth panel) is a profile from the quiet Sun. Figure adapted from \cite{2022arXiv221105333P}. Reproduced with permission.} \label{fig:mgiiintro}
\end{figure}

The flare chromosphere has been studied extensively using optically thick lines, which while presenting challenges with respect to extracting useful information due to their complex formation properties, offer important diagnostics of how the plasma responds to flares. Most notably, the H$\alpha$ line has been both observed and modelled in flare studies too numerous to exhaustively list or describe in detail here. Some examples include: \cite{1980ApJ...239.1036C},\cite{1984ApJ...282..296C}, \cite{1987ApJ...322..999C},\cite{1991ApJ...367..671C}, \cite{1991SoPh..135...65H}, \cite{1991ApJ...380..660G}, \cite{1992A&A...266..573G}, \cite{1993SoPh..143..141G}, \cite{2006SoPh..235..107L}, \cite{2015ApJ...813..125K}, \cite{2015ApJ...804...56R}. Some important numerical results inform us about how conditions in the upper chromosphere result in varying characteristics of the line, such as the depth of central reversal, are related to plasma conditions. The coronal pressure, for example, was found to be an important factor in determining the depth of the central reversal, with a high coronal pressure ($>100$~dyne cm$^{-2}$) required to markedly reduce the depth or fill in the reversal \citep[e.g.][]{1984ApJ...282..296C}. The pressure is closely related to density at the formation region of H$\alpha$, such that increasing the pressure forces the transition region, and hence H$\alpha$ formation height, deeper. As we will see in this section the \ion{Mg}{ii} central reversal depth is also seemingly related to conditions in the upper chromosphere. Additionally, exploring line widths and intensities could help constrain densities (via Stark wings) and temperatures (relative intensities of lines). Other lines have received significant attention in the past have included the \ion{Ca}{ii} H \& K resonance lines, and the infrared triplet \citep[e.g.][]{1975SoPh...42..395M,1992PASJ...44...63F,2002A&A...387..678F,Kuridze_2018,1999A&A...351..368D}. Non-thermal processes have also been studied numerically using chromospheric spectral lines. Non-thermal collisions between the particle beam and hydrogen or calcium have been shown to be a significant process, affecting the intensity and shape the lines \citep[e.g.][]{1993A&A...274..917F,2018A&A...610A..68D,2009A&A...499..923K}, and charge exchange between ion beams and hydrogen, producing highly redshifted Lyman line emission, has been suggested as a means to diagnose the non-thermal proton distributions in the lower atmosphere \citep[e.g.][]{1976ApJ...208..618O,1985ApJ...295..275C}.

In this section I introduce the \ion{Mg}{ii} NUV spectra as observed by IRIS, which being one of the strongest lines in the IRIS passbands has become a workhorse for studying the chromosphere, including during flares. Like the spectral lines of hydrogen and calcium, their characteristics are very sensitive to atmospheric properties, with various flare effects changing both the formation location as well as local plasma conditions. Magnesium is 18 times more abundant than Calcium, consequently forming higher in altitude and sampling the upper chromosphere, a useful region for understanding energy deposition during flares. The \ion{Mg}{ii} h \& k Doppler width is much smaller than that of H$\alpha$, offering advantages in sensitivity to both Doppler motions and turbulence.

\subsubsection{Observational Characteristics of \ion{Mg}{ii} Emission}
The \ion{Mg}{ii} NUV spectra, comprising the h \& k resonance lines ($\lambda2803.52$~\AA\ \& $\lambda2796.34$~\AA, respectively), the subordinate triplet ($\lambda2791.60$~\AA, $\lambda2798.75$~\AA\ \& $\lambda2798.82$~\AA, the latter two of which are blended), and who's lower energy levels are the upper levels of the resonance lines, and the quasi-continuum that lies between them, offer a wealth of information about the chromosphere. As one of the strongest and most commonly observed lines in the IRIS dataset, they have been well studied both observationally and in models. They are, however, somewhat of a menace to interpret, requiring complicated radiation transfer modelling including partial frequency redistribution (PRD; meaning there is a coherency between incident and scattered photons, which effects conversion of photons from core to wing) to help extract the information they carry. Obtaining a strong almost one-to-one match even in the quiet Sun still proves very challenging, likely due to both the complexity of the radiation transfer involved, and the assumed model atmospheres (the main chronic problem is the line width, which is much too narrow in simulations). While we make progress in obtaining more consistent model-data comparisons we learn more about the formation properties of the lines and the flaring conditions that we can infer about the plasma. 

The \ion{Mg}{ii} lines were comprehensively studied in the quiescent chromosphere most recently by \cite{2013ApJ...772...89L,2013ApJ...772...90L} and \cite{2013ApJ...778..143P}. While in active regions and flares their formation properties likely deviate from the description that follows, the quiescent studies form a basis for understanding these strong lines. These lines form throughout the chromosphere, with cores forming in the upper chromosphere, and wings forming from the upper photosphere through mid chromosphere. The resonance lines appear centrally reversed in most quiet Sun conditions (sunspots being the notable exception, though there the subordinate lines remain in absorption), with the core flanked by two emission peaks. The core is referred to as the k3 (or h3) component, and the emission peaks are collectively the k2 (or h2) components, with the blue peak referred to as k2v (or h2v) and the red peak as k2r (or h2r). Figure~\ref{fig:mgiiintro} shows both a quiet sun and flare \ion{Mg}{ii} profile to illustrate these features. This central reversal forms because the line source function is decoupled from the Planck function (that is, the local temperature), and falls with increasing altitude, so intensity at the height at which optical depth is unity ($\tau_{\lambda} = 1$; the surface from which we see the emergent intensity at some $\lambda$) is smaller than the intensity of the emission peaks, which have a slightly smaller opacity and form somewhat deeper. Their width, the asymmetry of the strength of the flanking peaks, their intensity, the depth of the reversal, and the k/h intensity ratio all show variations depending on the source conditions \citep[e.g][]{1973A&A....22...61L,1976ApJ...205..599K,1981A&A...103..160L}. The k/h intensity ratio, $R_{k:h}$\footnote{This is typically the integrated intensity, but recently \cite{2022ApJ...926..223Z} showed in their study of the \ion{Si}{iv} resonance lines it can be instructive to consider the ratio as a function of wavelength.}, has a typical value around $R_{k:h}=1.2$, indicative of optically thick line formation ($R_{k:h}=2$, the ratio of their oscillator strengths, in optically thin formation conditions). The subordinate lines are generally in absorption, unless there is additional heating at large column depth where they typically form \citep[e.g][]{2015ApJ...806...14P}. Note that modelling suggests that in the flaring case, the subordinate lines form much higher in altitude, and so subordinate line emission in flares is likely not a signature of deep heating, instead representing a heated mid-upper chromosphere \cite[see, e.g.][]{2019ApJ...871...23K,2019ApJ...879...19Z}

In flares the \ion{Mg}{ii} h \& k profiles are seen to significantly increase their intensity (by several factors to greater than an order of magnitude), broaden (with $FWHM>1$~\AA, compared to $FWHM<\sim0.5$~\AA\ pre-flare), exhibit redshifted cores (several tens of km~s$^{-1}$) and/or extreme red wing asymmetries, and to fill in their central reversal, appearing single peaked or with only a very shallow reversal \citep[e.g.][]{2015A&A...582A..50K,2015SoPh..290.3525L,2018ApJ...861...62P}. These lines can appear rather Lorentzian in shape in many flare spectra. The subordinate lines come into emission and display many of the same characteristics as the resonance lines. In some cases, blue wing asymmetries are observed \citep[e.g.][]{2015A&A...582A..50K,2018PASJ...70..100T,2019ApJ...878L..15H}. The k/h line intensity ratios during flares still indicate optically thick emission, and have been reported to decrease slightly. \cite{2015A&A...582A..50K} found $R_{k:h}=1.07-1.19$ in an M-class flare, and \cite{2018ApJ...861...62P} found an average of $R_{k:h} = 1.16$ from their larger survey. The range of observed $R_{k:h}$ values seems smaller in the flaring region \citep[][]{2015A&A...582A..50K}. Finally, \cite{2016ApJ...819...89X} and \cite{2018ApJ...861...62P} found that profiles located at the leading edge of some flare ribbons appeared very different to the profiles located in the bright ribbon segments. They contained deep central reversals, were much broader, had slightly blueshifted cores, and asymmetric emission peaks. The \ion{Mg}{ii} profiles from flares can vary on short timescales and small spatial scales (sometimes frame-to-frame, and pixel-by-pixel), suggesting they are extremely sensitive to plasma conditions, and therefore flare energy input.

\subsubsection{Flare Modelling of \ion{Mg}{ii}}
Efforts to model the \ion{Mg}{ii} spectra with electron beam driven flare simulations generally leads to profiles that behave qualitatively as we might expect, but contain important quantitative issues \citep[e.g.][]{2015SoPh..290.3525L,2017PhDT.......423K,2016ApJ...827..101K,2016ApJ...827...38R,2019ApJ...883...57K,2019ApJ...885..119K}. For example results from \radyn\ + \rh\ or using semi-empirical flare atmospheres shows that the \ion{Mg}{ii} spectra have an intensity increase (but are usually \textsl{too} intense, by up to approximately an order of magnitude or more), have redshifts and red-wing asymmetries (but the occasionally observed blue-wing asymmetries are harder to explain in the models), are broadened (but are significantly more narrow in the mid-far wings than observations, with observations in the range $FWMH\sim[0.5-2]$~\AA\ but in typical modelling $FWHM<0.5$~\AA), have subordinate lines in emission (but which are also too narrow and too weak relative to the resonance lines, by up to several factors), and have shallower central reversals (but it is difficult to synthesise the single peaked spectra that are observed).  Understanding the source of these differences can lead us to better understanding of plasma conditions and how to improve our modelling efforts to obtain those conditions. 

In a data-driven study of the 2014-March-29th X-class flare \cite{2016ApJ...827...38R} analysed \ion{Mg}{ii} k line observations, comparing them to forward modelling using \radyn\ + \rh. RHESSI hard X-ray observations were used to obtain the non-thermal electron distribution over time, using 2796~\AA\ IRIS SJI to estimate the energy fluxes (which ranged $F\sim[4\times10^{10}-10^{11}]$~erg~s$^{-1}$~cm$^{-2}$), and were split into 16 `threads', the timings of which were defined by the derivative of the GOES X-ray Sensor-B channel (XRS-B; 1-8\AA\ soft X-rays). Each individual peak in the soft X-ray derivative was proposed to represent the injection of particles into the chromosphere, and the duration of the heating phase of each thread was taken from the duration of each soft X-ray peak. These threads were individually processed through \rh\ to synthesise \ion{Mg}{ii} spectra, and were subsequently averaged in time to mimic the contribution from multiple threads over the flaring area (the heating and relaxation times of some threads overlapped). IRIS spectra were averaged over the source region of hard X-rays, and compared to the thread-averaged synthetic spectra. As hinted above, this comparison was less than ideal, with synthetic \ion{Mg}{ii} spectra having central reversals in the cores, that were much too narrow, and which at times exhibited blueshifts. There were some qualitative matches, however, with strong intensity enhancements and downflows producing redshifted line cores of up to a few tens of km~s$^{-1}$. Contrary to what is suggested by \cite{2016ApJ...827...38R}, I think that the presence of a strong k2v peak in the synthetic spectra is indicative of a strong \textsl{downflow} in the upper chromosphere in the model, rather than upflows. In their Figure 13, it can be seen that the line core (the centrally reversed part of the line) is redshifted, indicating a downflow. This would shift the absorption profile to the red, meaning that k2r photons from the red peak are more strongly absorbed than k2v blue peak photons that can more easily escape. The result is a brighter blue peak relative to red peak, producing an asymmetry. The effect of mass flows on the absorption of emission in optically thick lines has been discussion in detail in the context of \ion{Ca}{ii}, H$\alpha$, and \ion{Mg}{ii} in both acoustic shocks and flares: \cite{1992ApJ...397L..59C,2015ApJ...813..125K,2016ApJ...827..101K}. 

To address the sources of these model-data discrepancies \cite{2016ApJ...827...38R} studied the line formation properties and manually varied a snapshot of the \radyn\ atmosphere input to \rh, introducing microturbulence. They found that introducing a large microturbulent velocity ($v_{turb} = 27$~km~s$^{-1}$, compared to the $v_{turb} = 10$~km~s$^{-1}$ assumed in the original model) could broaden the line core, but could not model the extended wings. A similar conclusion was reached by \cite{2019ApJ...878L..15H}, who performed a model-data study comparing a flare jointly observed by IRIS and Big Bear Solar Observatory/Goode Solar Telescope \citep[BBSO/GST;][]{2012ASPC..463..357G} to an F-CHROMA \radyn\ model \footnote{\url{https://star.pst.qub.ac.uk/wiki/public/solarmodels/start.html}} with inputs most closely aligned with non-thermal electron distribution parameters discerned from RHESSI hard X-rays. They processed snapshots of that simulation through \rhpar, with different values of  $v_{turb} = [10,20,30]$~km~s$^{-1}$, and while it seems that a single value of turbulent velocity was not able to appropriately model the line, a weighted combination was more successful at capturing the width at the time a blue-wing asymmetry appeared, which was the main focus of that work. A means to estimate the actual turbulent velocity in flares that contributes to line broadening is to measure the non-thermal line width of an optically thin line (or ideally multiple lines), which does not suffer from opacity broadening effects that muddy the waters. There are not many strong optically thin lines originating in the chromosphere so obtaining this value at multiple formation temperatures is difficult, but even a rough guide would be very useful. The \ion{O}{i} 1355.6~\AA\ line observed by IRIS is optically thin in the quiet Sun \citep[][]{2015ApJ...813...34L}, and preliminary modelling results suggests that it remains so during flares \citep[e.g.][ plus Prof. M. Carlsson \textsl{private communication 2022}]{2019ApJ...871...23K}. This line has been relatively little studied observationally, but some estimates of $v_{nthm} \sim 9-10$~km~s$^{-1}$ were obtained in an M class flare \cite{2017PhDT.......423K}, and similar values have been seen in C-class flares (Dr. Sargam Mulay, \textsl{private communication 2022}), a slight increase from the $\sim7$~km~s$^{-1}$ measured in plage \citep[][]{2015ApJ...809L..30C}. More flare observations of this important line should be studied, especially in relation to \ion{Mg}{ii} line widths. The value required by \cite{2016ApJ...827...38R} is rather high compared to the (albeit, limited) estimates courtesy of \ion{O}{i} 1355.6, and could be unfeasibly large, approaching the sound speed of the chromosphere. Though, flares are very complex environments and it remains to be seen what a full survey of \ion{O}{i} reveals, and understanding how much of the missing width is due to turbulence is important to constrain the requirements of other sources to explain the deficit.

Returning to  \cite{2016ApJ...827...38R}, they also experimented with manually raising the electron density by a factor of 10 (to $n_{e}>10^{14}$~cm$^{-3}$) in a narrow region at the base of the transition region (though not in a self-consistent manner so that the temperature and other properties were fixed), which had the effect of filling in the central reversal. This was because the larger density allowed a greater degree of collisional coupling to local conditions\footnote{The line source function is, roughly, a measure of the ratio of the upper and lower level populations, which depends on local conditions via collisions, and non-local conditions via the radiation field. The dominance of radiative processes over collisions, and the escape of radiation near $\tau_{\lambda} = 1$, means that the source function deviates from the Planck function. In fact, this is what creates the central reversal in optically thick lines. While the Planck function increases with height due to larger temperatures, the source function decreases as it deviates from the Planck function. The line core source function is smaller at the height where the line core k3 photons can escape ($z(\tau_{\lambda,core} = 1)$) than the source function at a deeper altitudes where the photons that create the flanking k2 peaks escape ($z(\tau_{\lambda,k3} = 1) > z(\tau_{\lambda,k2} = 1)$). Increasing collisions (e.g. via a larger density and temperature) helps to maintain or increase the population of the upper level via collisional transitions, and the source function more closely tracks the local Planck function, reducing the difference between the k2 and k3 source functions, such that the central reversal can reduce in depth or disappear.}, and the \ion{Mg}{ii} k line core source function increased with height, tracking the rise of the Planck function more closely, past the point at which the $\tau_{\lambda} = 1$. This suggests that in actual flares, extremely large densities are present in the upper chromosphere/lower transition region, to drive the line to appearing single peaked. 

\begin{figure}[h]
\begin{center}
\includegraphics[width=\textwidth]{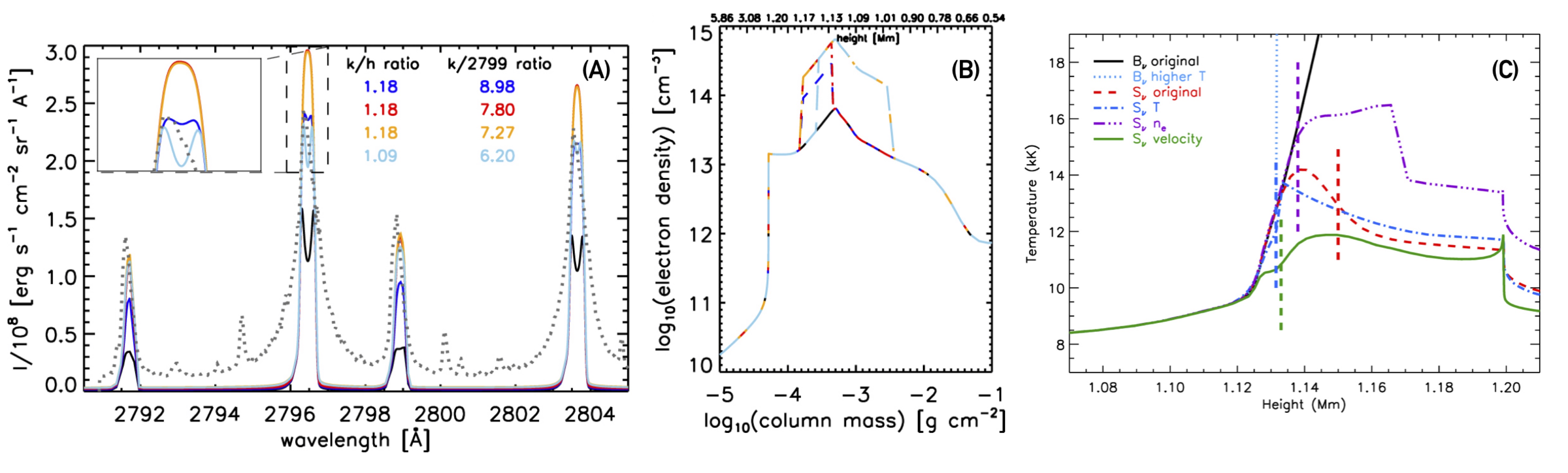}% This is a *.eps file
\end{center}
\caption{Illustrating the effect of increased electron density in the upper chromosphere on the \ion{Mg}{ii} NUV spectrum. Panel \textbf{(A)} shows the \ion{Mg}{ii} spectra synthesised from a \radyn\ simulation, using \rh, where each colour represents a different modification to the electron density, the stratification of which is shown in panel \textbf{(B)}. The black line is the original, and the dotted line is a sample observed flare spectrum from IRIS. The resonance to subordinate line ratios are indicated. An order of magnitude increase in the electron density to $>5\times10^{14}$~cm$^{-3}$ is required to produce a single peaked profile. Panel \textbf{(C)} shows that increasing the electron density (purple line, compared to all others) results in a stronger coupling to the Planck function (black line), hence the lack of central reversal in that instance. The vertical lines in that panel are the formation heights of the \ion{Mg}{ii} k line core for each model atmosphere. Figure adapted from \cite{2017ApJ...842...82R}. \copyright AAS. Reproduced with permission.} \label{fig:mgiisinglepeak}
\end{figure}

Given the discrepancies identified by \cite{2016ApJ...827...38R} and other authors, \cite{2017ApJ...842...82R} decided to perform a much larger parametric study to determine what aspects of the flaring atmosphere had to change in order to produce \ion{Mg}{ii} NUV spectra more consistent with observations. They took a snapshot from one of the \radyn\ simulations from \cite{2016ApJ...827...38R}, and manually varied the temperature, electron density, and velocity structure of the flaring chromosphere in a systematic way, that were processed through \rh. They varied one parameter at a time, and did not re-solve the atmosphere to hydrostatic equilibrium given the updated parameter, in order to discern the direct impact of, for example, temperature variations by itself. This means that charge was not conserved in their models, and the atmospheres investigated were not self-consistent. Still, \cite{2017ApJ...842...82R} provided great insights into plasma conditions that could be producing the observed profiles.

Introducing a steeper temperature rise in the upper chromosphere through the transition region \footnote{Temperatures originally climbed from from $\log_{10} T \sim 4.1$~[K]  to $\log_{10} T \sim 4.6$~[K] between column mass of $\log_{10} \sim [-3.6, -4.3]$~[g cm$^{-3}]$, before then rapidly attaining transition region / coronal temperatures a column mass of $\log_{10} > -4.4$~[g cm$^{-3}$]. They were modified to instead increase from $\log_{10} T \sim 4.1$~[K] to $\log_{10} T \sim 4.9$~[K] at a column mass of $\log_{10} \sim [-3.4, -3.6]$~[g cm$^{-3}]$, before more gradually increasing to $\log_{10} T > 5.5$~[K]  between column mass of $\log_{10} = [-3.6, -4.3]$~[g cm$^{-3}$].}, the formation region of the \ion{Mg}{ii} h \& k line cores, led to weaker (by a factor $\sim1.5-2$), single peaked profiles, but did not sufficiently enhance the subordinate lines so that the k:subordinate line ratio was much too high compared to the observations ($R_{k:sub} \sim15$ compared to the observed $R_{k:sub} \sim4$). Here, introducing higher temperatures forces \ion{Mg}{ii} to a deeper formation region due to thermal ionisation. At those deeper altitudes, there is a greater average electron density where the line core forms, thus stronger coupling to the local conditions that acts to fill in the central reversal. Increasing the mid chromosphere temperature did not lead to single peaked profiles but did decrease the k:subordinate line ratio closer to the observed values ($R_{k:sub} \sim8$). Adding temperature spikes of several thousand kelvin in the lower atmosphere (peaking at a column mass  $\log_{10} \sim -2.2$~[g cm$^{-3}$]) resulted in prominent spikes in the blue wings of both resonance and subordinate lines that are not observed (presumably the redshifted absorption profile results in the absorption of a similar feature in the red wing). 
 
Enhancing the electron density by a factor of ten through the formation region of the resonance line cores\footnote{Roughly, increasing the electron density by an order of magnitude though column masses of $\log_{10} \sim [-3.6, -3.8]$~[g cm$^{-3}]$, peaking at $ \log_{10} n_{e} \sim 14.8$ [cm$^{-3}$].} produced single peaked profiles, and actually drove the k:subordinate line ratio closer to observations ($R_{k:sub} \sim8$, so within a factor two of the observations), though also increased the line intensity by around a factor of two. This is illustrated in in the lefthand side of Figure~\ref{fig:mgiisinglepeak}, which shows the \ion{Mg}{ii} NUV spectra for different electron density stratifications with fixed temperature stratification. Here the enhanced electron density resulted in stronger collisional coupling to the local temperature, that is the Planck function, which can be seen the righthand panel of Figure~\ref{fig:mgiisinglepeak}. Raising the electron density below the core formation heights also drove the k:subordinate ratio lower, but did not produce single peaked resonance lines. In that scenario, the electron density increases the coupling of the subordinate lines to local conditions so that the source functions, and ultimately the emergent intensities, were larger. Seemingly, increasing the electron density deeper into the atmosphere affects the subordinate lines whereas the resonance lines largely require an increased upper chromospheric density. 

Varying the temperature and and electron density independently could not recover the very broad line wings. Instead, experiments with introducing extremely large downflows of $v\sim200$~km~s$^{-1}$ were attempted, in concert with weaker upflows $v\sim75-100$~km~s$^{-1}$. Combining unresolved flows did produce very broad resonance lines, but also too-weak subordinate lines. The blending of the far wings of the resonance lines with the quasi-continuum between them, and with the subordinate lines, was not well captured in those static models. While there is \textsl{prima facie} evidence from the modelling work of \cite{2017ApJ...842...82R} that unresolved bi-directional flows can broaden the lines, my own opinion is that extreme macrovelocity `smearing' of the lines is not the source of the missing widths far into the line wings. Such extreme flows are a difficult proposition. Downflows are typically modelled (and inferred from observations) as being much more modest, on the order of $v\sim$ a few$\times10$ km~s$^{-1}$, and concentrated in narrow, dense condensations. While complex flows often form in loop models, even sometimes with transient downflows of $v\sim150$~km~s$^{-1}$ in extended transition regions \citep[e.g.][]{2019ApJ...879...19Z}, by the time the condensations reach the chromosphere they have cooled, accrued mass and decelerated to be  $v<100$~km~s$^{-1}$ (more often slower, to a few$\times 10$ km~s$^{-1}$). That is not to say that extreme bi-directional flows are not what is happening in the actual chromosphere, but we therefore must determine a means to produce such large downflows in simulations that can capture the complex interplay between flows, the subsequent accrual and evacuation of mass, and the associated changes to opacity. One observational sanity check here could be to determine the k:h line ratio as a function of wavelength over the line, which may help determine the relative opacity of the secondary blueshifted component, which formed higher in the atmosphere. 

To summarise, the experiments of \cite{2017ApJ...842...82R} suggest that increasing the upper chromosphere temperature pushes the formation height of the \ion{Mg}{ii} lines deeper, but that perhaps we are missing a temperature increase through the chromosphere to enhance the subordinate lines. The likely culprit behind filling in the central reversal is an enhanced electron density in the upper chromosphere, perhaps (in my view: certainly!) related to the dense condensations produced in RHD models and studied extensively by Prof. Kowalski and collaborators. While extreme bi-directional velocity flows that are unresolved within an IRIS pixel appear to produce some of the missing line widths in their modelling, I am a bit more sceptical that these conditions can appear in the actual chromosphere.  \cite{2017ApJ...842...82R} clearly indicate that unresolved flows do play some role, but the flow magnitude required has not, to my knowledge, been inferred from typical observations of chromospheric spectral lines. A key focus for future flare modelling should be to (1) self-consistently combine several aspects of these important findings, e.g. a temperature rise through the lower-mid chromosphere would also raise the electron density and likely generate flows, and (2) to produce a flare model driven by some energy input that naturally produces the plasma conditions required by \cite{2017ApJ...842...82R}. We must also assess if the conditions that produce a closer match to \ion{Mg}{ii} observations do not produce conditions that results in a discordant match to other spectral lines (e.g H$\alpha$, \ion{Ca}{ii} 8542 or \ion{Ca}{ii} H \& K).

\begin{figure}[h]
\begin{center}
\includegraphics[width=\textwidth]{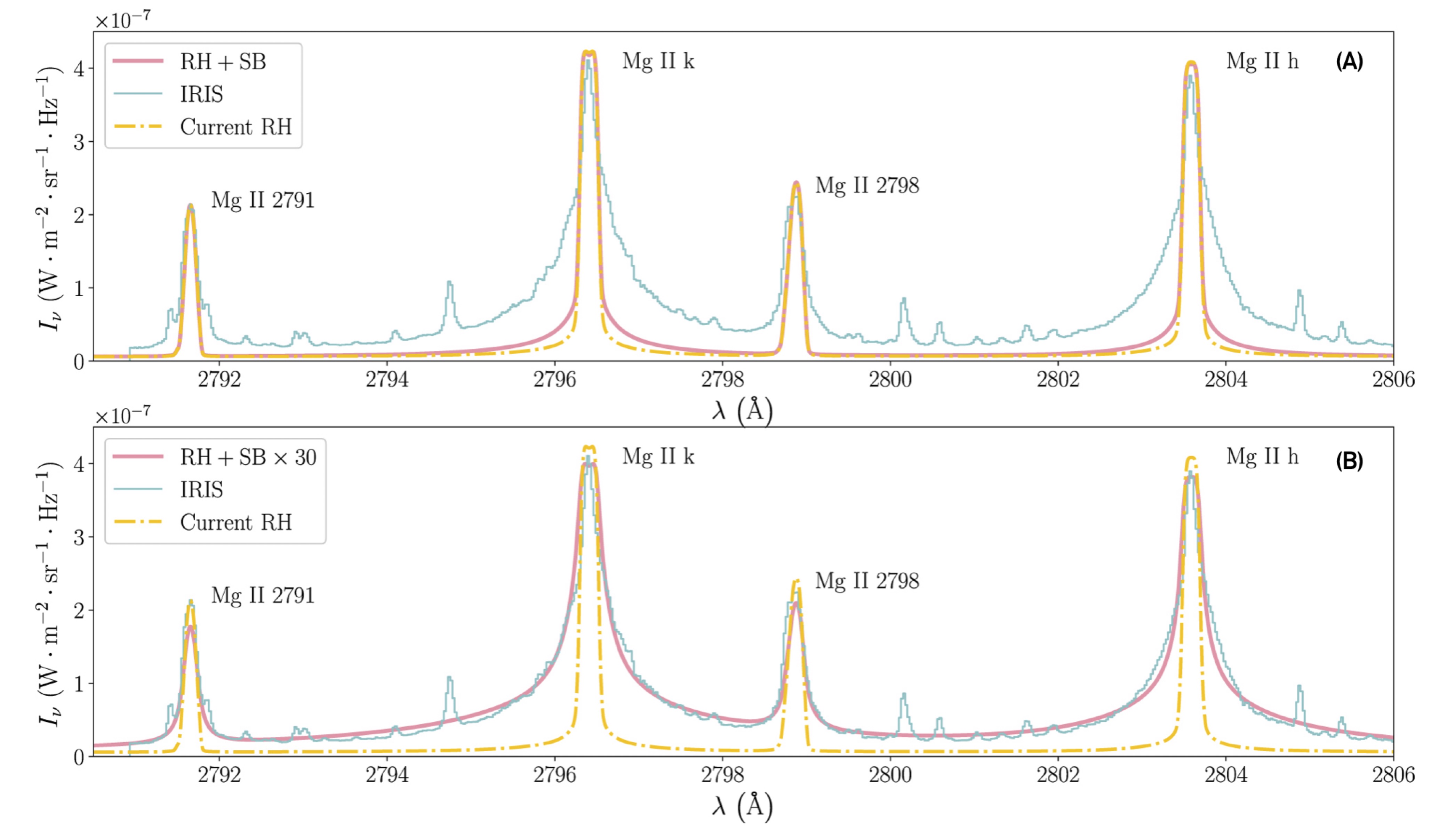}% This is a *.eps file
\end{center}
\caption{Improved treatment of Stark broadening for \ion{Mg}{ii} lines results in broader line wings. Panel \textbf{(A)} compares a \radyn\ simulation processed through \rh, where yellow dot-dashed is the synthetic \ion{Mg}{ii} line with standard Stark broadening, and the salmon coloured line is the line with improved Stark broadening of \cite{2019ApJ...879...19Z}. The blue line is the observation, scaled in intensity. Panel \textbf{(B)} illustrates that even with this improvement a further factor of $\times30$ Stark broadening would be required to produce a width consistent with the observation, indicating that we are still missing something in our models. Figure adapted from \cite{2019ApJ...879...19Z}. \copyright AAS. Reproduced with permission.} \label{fig:mgiiwidth}
\end{figure}

One alternative potential resolution to the question of the missing line widths in the models is that we have been underestimating quadratic Stark broadening (electron pressure broadening). Though not terribly important for \ion{Mg}{ii} spectral lines in the quiet Sun, the many orders of magnitude enhancement of the electron density in flares could result in pressure broadening playing an outsized role\footnote{Note that this is in contrast to the lower density flaring corona and transition region. \cite{2011ApJ...740...70M} found that pressure broadening played a negligible role in the broadening of optically thin lines in the corona and transition region.}, due to the fact that quadratic Stark broadening is a function of the electron density. In \rh\ the quadratic Stark effect is typically computed following classical Impact Theory with various approximations \citep[see discussion in ][]{2019ApJ...879...19Z}, including the adiabatic approximation. As \cite{2019ApJ...879...19Z} demonstrate, the adiabatic approximation is likely not valid for \ion{Mg}{ii}. Instead, impact-semiclassical-theory provides a better estimate, which is included in the STARK-B database\footnote{A database of Stark widths for various atoms/ions: \url{http://stark-b.obspm.fr/}}. \cite{2019ApJ...879...19Z} modified \rh\ to model \ion{Mg}{ii} Stark widths based on the STARK-B database, where the Stark width is a polynomial function of temperature and density. At the temperatures relevant for \ion{Mg}{ii} the STARK-B results have an order of magnitude greater value than is typically modelled by \rh. A \radyn\ simulation was produced with a very high peak electron beam flux of $F_{peak} = 5\times10^{11}$~erg~s$^{-1}$~cm$^{-2}$, that was ramped up and down with FWHM of $20$~s. Snapshots were processed through \rh\ with and without the improved Stark broadening. The inclusion of increased Stark broadening resulted in broader profiles compared to the typical flare loop models (though only the h \& k lines were strongly affected). Still, however, they remained too narrow compared to observations, and through experimentation it was found that a factor $\times30$ additional stark broadening was required over and above the STARK-B estimates to sufficiently broaden the lines. In that case, lines and quasi-continuum between the lines, were well reproduced, albeit with a factor $\sim36$ too high an intensity (which was the case with and without improved treatments of Stark broadening\footnote{This intensity discrepancy is likely a combination of the atmospheric stratification, and other factors such as 3D radiation transfer and the assumed filling factors of the observations. While this is a large intensity offset, such exploratory studies of chromospheric observables still teach us about important physical processes, even if we do not yet simultaneously solve both the problem of intensity and broadening here.}). Figure~\ref{fig:mgiiwidth} shows the results of improving Stark broadening, and that an additional broadening factor is still required to match the far wings. \cite{2019ApJ...879...19Z} followed \cite{2017ApJ...842...82R} by experimenting with different microturbulence stratifications but found that if the h \& k lines were suitably broadened via microturbulence, the subordinate lines had the wrong shape.

Additionally, single-peaked profiles were produced naturally by their model at several times. A detailed examination of the formation properties revealed that this happened when the electron density in the line formation region was exceptionally high (a result of the merging of several compressive flows), on the order of $n_e \simeq 8\times10^{14}$~cm$^{-3}$. The formation region was very narrow ($\Delta z = 32$~m, compared to $\Delta z \sim 100$s~m at other times), with a constantly increasing temperature. These results confirmed the findings of \cite{2017ApJ...842...82R} that the electron density in the formation region is a key factor in understanding the typically observed \ion{Mg}{ii} lines. \cite{2019ApJ...879...19Z} also note that unresolved flows in their simulations did broaden lines, but not to the extent required as the flows had slowed to $<50$~km~s$^{-1}$ in the chromosphere, and to $<10$~km~s$^{-1}$ at the time of single peaked profiles. They also had difficulty producing very asymmetric red wings, instead producing transient secondary components when the flows were still strong. It could be case that smearing over an exposure time typical of IRIS observations (up to a few seconds), and to IRIS resolution, would merge the shifted and stationary components into a more asymmetric type profile. Nevertheless, \cite{2019ApJ...879...19Z} successfully produced a single peaked profiles, and made progress towards understanding the missing line widths. Questions remain: (1) how do we explain the factor $\times\sim30$ still required to model the far wings? Could this be addressed by temperature/electron density enhancements in the lower atmosphere above that which we currently model with electron beams? Since the far wings are Lorenztian this seems like a plausible line of investigation; (2) how can we produce such high densities in flares in which the non-thermal electron flux derived from hard X-rays appears to be $<5\times10^{11}$~erg~s$^{-1}$~cm$^{-2}$. Have we perhaps been underestimating the energy fluxes so far, either of electron beams alone, or non-thermal electrons in conjunction with other sources of energy (for example non-thermal ions, conductive fluxes, MHD waves)?; (3) what is the impact of the pre-flare atmosphere in producing such high densities and in the overall evolution in general? 

There have been a few reports of blue wing asymmetries that are concentrated early in the development of flare sources \citep[e.g.][]{2015A&A...582A..50K,2018PASJ...70..100T,2019ApJ...878L..15H}. A suggestion was put forward by \cite{2018PASJ...70..100T} that blue wing asymmetries at ribbon fronts were produced by gentle evaporation of cool, dense chromospheric material into the corona, ahead of a hot bubble of material. This cool material is heated and dissipates. They created a cloud model that could produce \ion{Mg}{ii} h line blue wing asymmetries, along with the peak asymmetries observed. To my knowledge this has not been modelled in detail using flare loop models. A similar, but seemingly more extreme phenomenon, is the appearance of \ion{Mg}{ii} profiles that have unique shapes that only appear in the leading edge of flare ribbons (so-called ribbon fronts), found by \cite{2016ApJ...819...89X} and \cite{2018ApJ...861...62P}. As mentioned above, these exhibit quite different properties to brighter flare sources (namely blueshifted line cores, deep central reversals, and very broad profiles). When present, these profiles are located along the leading edge of propagating flare ribbons, and thus represent the very initial stages of energy deposition.  Other ribbon front spectral behaviour includes the \textsl{dimming} of \ion{He}{i} 10830~\AA\ before it brightens during the main part of the ribbons \citep[e.g.][]{2016ApJ...819...89X}. \cite{2021ApJ...912..153K} demonstrated using \radyn\ flare modelling that this dimming is caused by the presence of non-thermal particles in the chromosphere, and that a weaker flux with a harder distribution (that is, greater proportion of high energy electrons than low energy electrons) resulted in stronger dimming that was sustained for slightly longer. However, observed ribbon front behaviour can persist for up to $120-180$~s (though they can also be shorter in duration), whereas the models of \cite{2021ApJ...912..153K} predicted only a few seconds. Once the chromosphere was hot enough in those simulations, the \ion{He}{i}~10830 line was driven into emission.  The causes of flare ribbon fronts (which do not appear appear uniformly along the ribbon), and how they transition to the more typical bright ribbons we have historically studied, is not known. Work is in-progress to address the ribbon front problem using \radyn\ modelling of electron beam driven flares: \cite{2022arXiv221105333P} investigates the relation between the magnitude of energy flux deposited and response of the \ion{Mg}{ii} ribbon front-like profiles, finding that weak energy fluxes are more consistent with ribbon-front like profiles and that stronger energy fluxes produce more `standard' ribbon profiles. That study used the same simulations from \cite{2021ApJ...912..153K}, and those simulations that were most consistent with \ion{He}{i} 10830~\AA\ ribbon front observations also resulted in \ion{Mg}{ii} spectra that were comparable to ribbon-front observations. The implication here is that there are potentially different stages of energy deposition to explain the evolution from ribbon-front to ribbon spectral profiles, and follow on from \cite{2022arXiv221105333P}, led by myself, is investigating how to obtain longer lived ribbon fronts in both \ion{He}{i} and \ion{Mg}{ii}. From these two on-going studies it is clear that the \ion{Mg}{ii} ribbon front profiles can strongly constrain the characteristics of initial energy deposition into the chromosphere, and that high spatial-, temporal-, and spectral resolution observations in other wavelengths should focus on ribbon leading edges. 

\subsection{Hot Flare Plasma Observed by IRIS}
Prior EUV observations of hot flare lines have shown anomalously broad lines, of unknown origin \citep[e.g.][]{2011ApJ...740...70M,2015SoPh..290.3399M}. While several suggestions have been made, a definitive solution remains elusive (as you have no doubt realised by now, line widths are a sore spot for flare modellers). As discussed in Paper 1 in relation to probing the duration and magnitudes of chromospheric evaporation,  the \ion{Fe}{xxi}~1354.1~\AA\ line observed by IRIS offers a window at high spatial, temporal, and spectral resolution on hot flare plasma at $\sim11$~MK. Here I discuss a few studies that attempted to model \ion{Fe}{xxi} 1354.1~\AA\ emission in flares, concentrating on line broadening. This is a different problem than the \ion{Mg}{ii} missing width, as opacity effects play no role for \ion{Fe}{xxi} 1354.1~\AA\ emission, which is a forbidden line.

\begin{figure}[h]
\begin{center}
\includegraphics[width=0.75\textwidth]{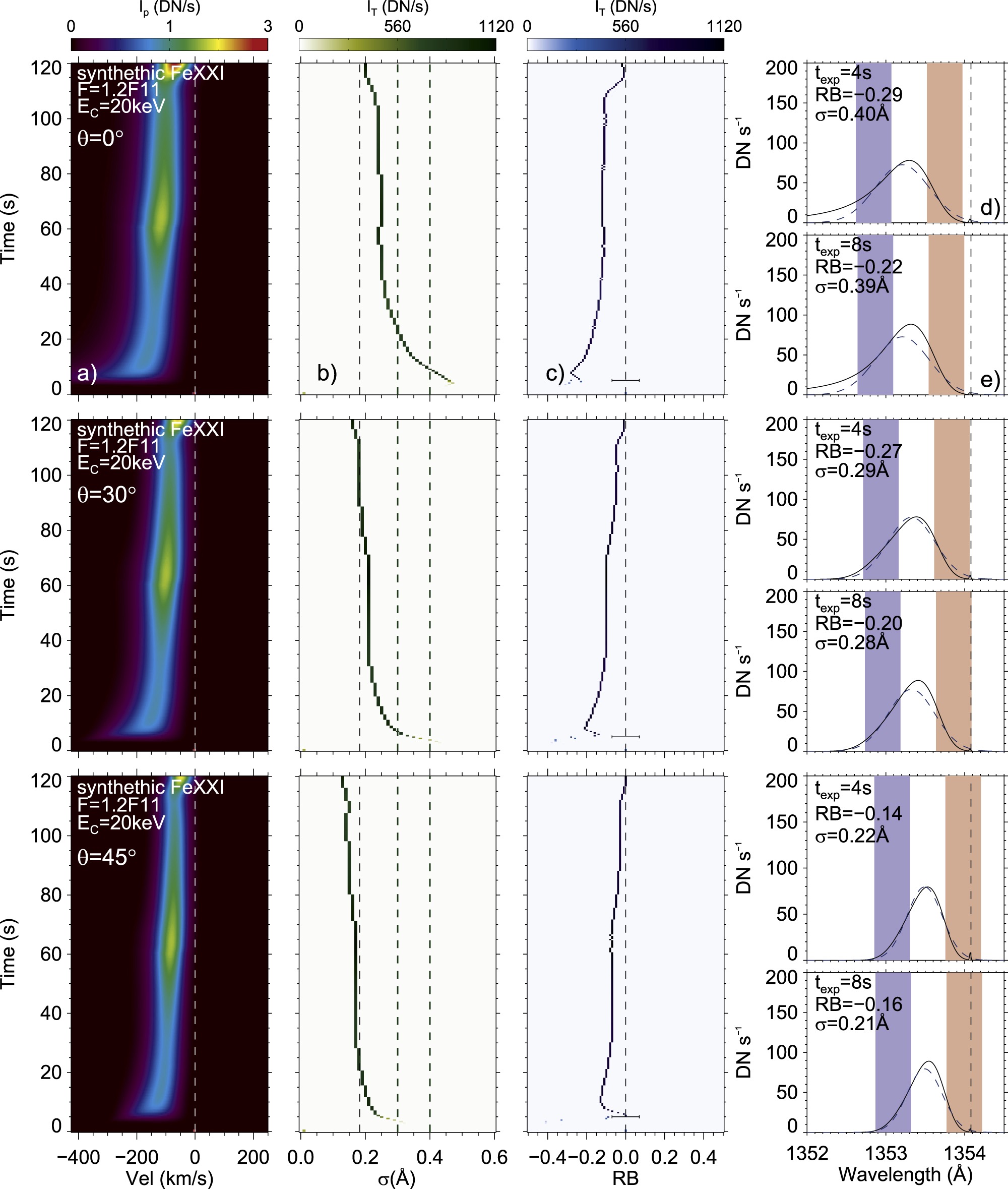}% This is a *.eps file
\end{center}
\caption{Synthetic \ion{Fe}{xxi} 1354.1~\AA\ line profiles from a \radyn\ simulation that modelled the superposition of loops to attempt to explain line broadening. Panel \textbf{(A)} shows the spectra as a function of time. Panel \textbf{(B)} is the line width as a function of time, where the vertical lines are the minimum width (leftmost) and typical ranges from observations (two rightmost). Panel \textbf{(C)} is the red-blue wing asymmetry as a function of time, where the horizontal line shows the typically observed values. Panels \textbf{(D-E)} are synthetic spectra showing the difference in assumed exposure times. The coloured bands represent the areas used to calculate the red-blue wing asymmetry and the dashed curve is a single Gaussian fit to the spectra. The vertical line is the rest wavelength. The remaining panels show the same, but for different inclination angles of the loop to the solar surface. Figure adapted from \cite{2019ApJ...879L..17P}. \copyright AAS. Reproduced with permission.} \label{fig:fexxiwidth}
\end{figure}

The \ion{Fe}{xxi} 1354.1~\AA\ line has been observed in numerous flares by IRIS \citep[e.g.][]{2015ApJ...807L..22G,2015ApJ...803...84P,2016ApJ...816...89P,2015ApJ...799..218Y,2015ApJ...811..139T}. They are observed to be largely symmetric, with significantly enhanced line widths. They are initially weak and broad, and become more narrow and intense over time. The line widths during flares have ranged from the instrumental + thermal width $W\sim0.43$~\AA\ (assuming ionisation equilibrium) at loop tops to $W\sim[0.5-1]$~\AA\ or larger in ribbons. \cite{2020ApJ...900...18K} performed a superposed epoch analysis similar to \cite{2015ApJ...807L..22G}'s Doppler shift analysis, to understand the typical evolution of \ion{Fe}{xxi} line widths over time in the 2014-September-10th X class flare. That event showed a large amount of scatter during the impulsive phase of the flare, but with $W\sim[0.6-1.2]$~\AA, peaking after $t\sim200$~s, before gradually returning to pre-flare values over the subsequent $500$~s or so.

 A popular suggestion for the origin of the broad line profiles of hot lines is the superposition of of flows along the line of sight from numerous Doppler shifted line components. Using the \radyn\ code, \cite{2019ApJ...879L..17P} explored this idea. They produced several field-aligned flare simulations, with a $t=60$~s heating duration. Synthetic \ion{Fe}{xxi} emission was produced, with Doppler shifts applied as appropriate as a function of height along the loop. From those simulations they constructed both single and multi-stranded loop models, which for the latter had 100 identical threads each with a randomly selected start time within $15$~s of the first thread start time. Other loop lengths and energy fluxes were also tested, but no change to the overall conclusions were found. The threads were then orientated in several ways that investigated the effects of loops being co-spatial or along a ribbon-like structure within 20 IRIS pixels, either aligned in the same angle or at different orientations. Emission from each set-up was summed to mimic different scenarios of IRIS looking through different lines of sight. \cite{2019ApJ...879L..17P} found that there was a non-negligible asymmetry, with an anti-correlation between broadening and asymmetry (broader lines were more asymmetric), in contrast to observations which showed largely symmetric lines no matter the width, in each of their scenarios. Narrow profiles were quite symmetric, and were largely due to superposition of several upflows that had similar magnitudes, but the superposition of loops was unable to characterise the broad, symmetric \ion{Fe}{xxi} profiles in this case. Figure~\ref{fig:fexxiwidth} shows a sample of these experiments, for a single loop model with different orientations. The spectra and resulting broadening are shown, where it is clear that asymmetries are present.

\begin{figure}[h]
\begin{center}
\includegraphics[width=0.75\textwidth]{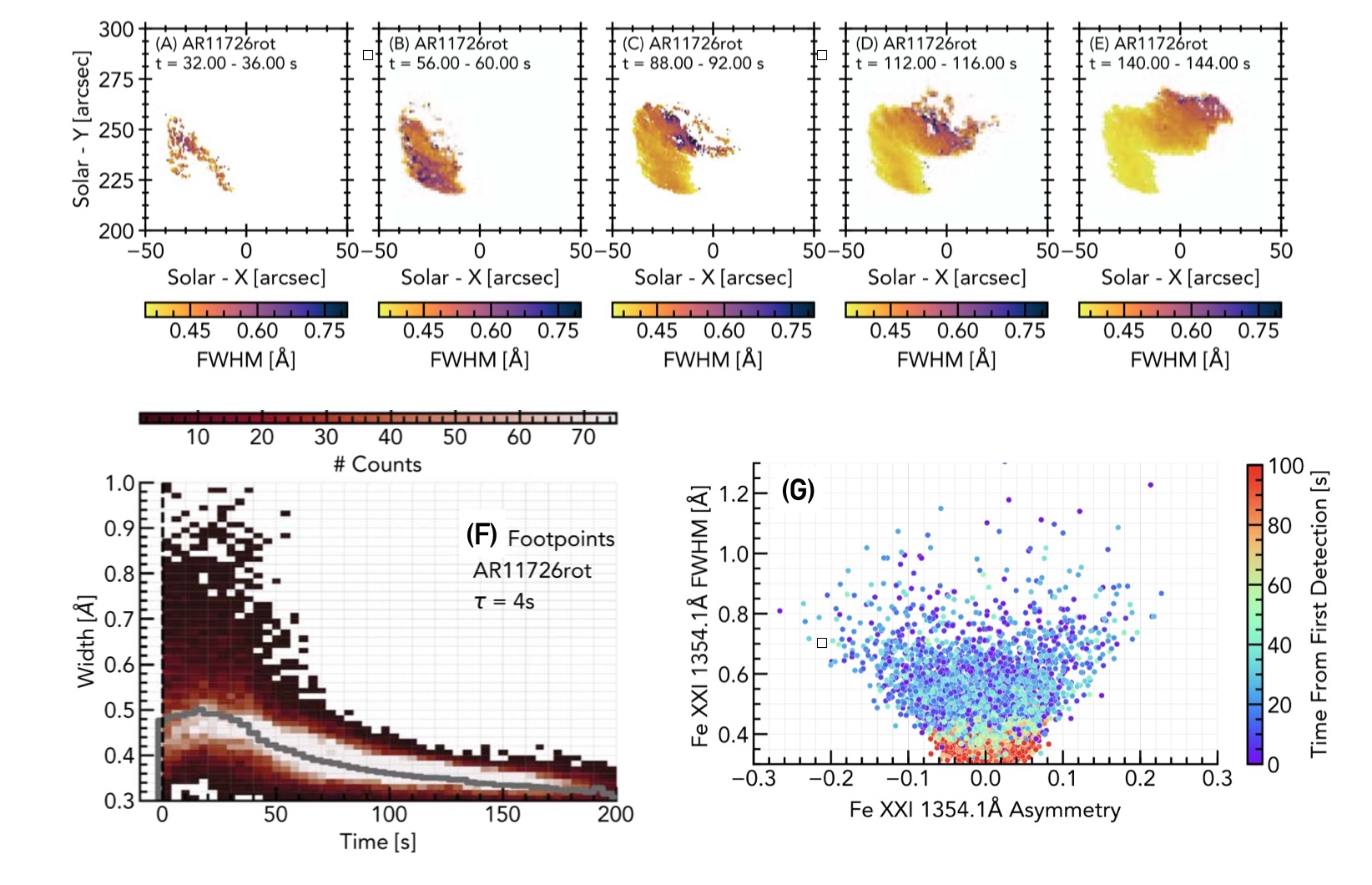}% This is a *.eps file
\end{center}
\caption{Evolution of \ion{Fe}{xxi} widths in a \radynarcade\ model. The top row (A-E) show maps of the line widths at various snapshots, illustrating that the footpoints and lower legs exhibit broader profiles. Taking all these pixels and producing a superposed epoch analysis (panel F) indicates that in comparison to an observations the line widths are too narrow. Panel (G) further illustrates that the profiles are too narrow, and that larger-than-observed asymmetries appear for some of the broadest profiles (colour represents elapsed time from the first moment that \ion{Fe}{xxi} was detected in that pixel). Figure adapted from \cite{2020ApJ...900...18K}. \copyright AAS. Reproduced with permission.} \label{fig:kerrfexxiwidth}
\end{figure}

\cite{2020ApJ...900...18K} also studied synthetic \ion{Fe}{xxi} line widths using \radyn\ simulations. The model framework they developed, \radynarcade, is described in Paper 1. From the field-aligned loops grafted onto the observed magnetic field skeleton, the superposition along the line of sight, and loop geometry was automatically taken into account. Qualitatively they produced synthetic \ion{Fe}{xxi} emission that largely followed the observations. There was a line broadening that was strongest at flare footpoints, and which narrowed along the loops towards looptops. However, while some profiles exceeded $W\sim0.8$~\AA. the majority of the \ion{Fe}{xxi} lines only reached $W\sim[0.5-0.6]$~\AA, much narrower than observed. There was not a very strong correlation between asymmetry and line width (broad profiles could be both symmetric or asymmetric), but the largest asymmetries were associated with the broadest profiles. The majority of the profiles were fit well with a single Gaussian, with only a subset requiring multiple components. There was also not a strong correlation between line width and Doppler shift, in contrast to some observations of hot flare lines studied by \cite{2011ApJ...740...70M}. Finally, a synthetic superposed epoch analysis showed again a qualitative similarity to observations, but with profiles that were too narrow (with synthetic FWHM$\sim[0.4-0.8]$~\AA, clustered around FWHM$\sim0.5$~\AA, compared to observations with a range of FWHM$\sim[0.4-1.4]$~\AA, clustered around FWHM$\sim0.8$~\AA), and profiles that both peaked and narrowed in width too quickly (in observations the profiles took several hundred seconds to narrow towards pre-flare values, compared to $t<100$~s in the model). Figure~\ref{fig:kerrfexxiwidth} shows a map of \ion{Fe}{xxi} from the \radynarcade\ model at various snapshots, alongside the synthetic superposed widths, and the asymmetries versus line widths, illustrating the discrepancies. These results agree with the earlier findings of \cite{2019ApJ...879L..17P}.

As set out nicely by \cite{2019ApJ...879L..17P}, there are several possible physical conditions that the \radyn\ modelling did not account for, which might explain how to obtain a closer match to the IRIS observations. Turbulence (including MHD wave turbulence) could broaden lines symmetrically. In fact, \cite{2022ApJ...931...60A} recently demonstrated that by suppressing thermal conduction in a \radyn\ simulation, via non-local effects or turbulence \citep[e.g.][]{2018ApJ...865...67E}, could lengthen the gradual phase of a flare, and produce a flow pattern more consistent with observations \citep[e.g.][]{2009ApJ...699..968M}. Taking the turbulent mean free path of the best fit model, \cite{2022ApJ...931...60A} were able to estimate the broadening associated with turbulence for numerous lines (including hot lines from high charge states of Fe), finding that they were broadened substantially and symmetrically. There are plans to perform follow on studies to the modelling of \cite{2020ApJ...900...18K}, using these \radyn\ updates, and with the observed line widths from IRIS \& \textsl{Hinode}/EUV Imaging Spectrograph (EIS) observations as constraints on the degree of suppression to include. Using \preft, Dr. William Ashfield and Dr. Dana Longcope are exploring the creation of MHD turbulence following loop retraction with added drag (\textsl{private communication  2022}). I eagerly await the application of their modelling to the formation of \ion{Fe}{xxi}. Another source of broadening could be non-equilibrium effects, such that the ion temperature is very much larger than the equilibrium value of $11.2$~MK. This would require an ion temperature on the order $40-60$~MK, which also likely requires decoupling of the ion and electron temperatures \citep[see also][]{1985SoPh...98..267D,2018ApJ...864...63P}. Given high densities in flare footpoints, it is not clear if such extreme non-equilibrium processes apply, but the \hydrad\ code is the ideal resource to study this in flares. Finally, Alfv\'enic waves propagating downwards from the magnetic reconnection site could broaden spectral lines via ion motions (see Section~\ref{sec:mechs}).

\subsection{The Transition Region Observed by IRIS}
The extreme gradients through the transition region make it an important interface for mass and energy transport during flares. Strong lines produced in the transition region that are observed by IRIS are the \ion{Si}{iv} and \ion{C}{ii} resonance lines. Aside from study of their Doppler shifts, \ion{C}{ii} has been relatively little studied in flare loop models. \ion{Si}{iv} 1394~\AA\ and 1402~\AA, however, have been modelled in a few studies. I discuss their Doppler shifts in Paper 1, but here focus on their intensity ratio, and what that might tell us about temperatures and densities in the flaring transition region. 

These lines have been used to infer flows in flares from their Doppler shifts, under the assumption that they are optically thin. They increase in intensity, broaden to a line width of similar magnitude to \ion{Mg}{ii}, and exhibit red wing asymmetries indicative of mass flows on the order of a few$\times10$ to $100$~km~s$^{-1}$ \citep[e.g.][]{2015ApJ...811..139T,2017ApJ...848..118L,2020ApJ...896..154Y}. Non-Gaussian line shapes have been observed in flare ribbons, but most observations suggest optically thin formation from the ratio of the 1394~\AA\ / 1402~\AA\ intensities ($R_{1394/1402} = 2$; though it is often the case that only one of the lines is included in the IRIS lineslists), with exceptions discussed later in this section. Given the optically thin assumption, most flare modelling of this line was performed by computing the emissivity from CHIANTI atomic data alongside the stratification of physical variables from flare loop model atmospheres, and summing through height in some fashion to obtain the total intensity. However, some quiescent Sun studies suggested that effects of non-equilibrium radiation transfer, photoexcitation, or charge exchange may be important in setting the \ion{Si} ion fraction stratification \citep[e.g.][]{2017ApJ...842...19D,2017A&A...603A..14D,2018A&A...610A..67D,2019ApJ...871...23K}. Observations of line ratios in stellar flares have also suggested that the resonance lines of \ion{Si}{iv}, and also of \ion{C}{iv} exhibit opacity effects \citep[e.g.][]{2002A&A...390..219B,1999A&A...351L..23M} and form under optically thick conditions, or, at least with some non-negligible optical depth $\tau>0.1$.

To determine the importance of radiation transfer effects on the formation of \ion{Si}{iv} resonance lines during flares, \cite{2019ApJ...871...23K} simulated a large number of electron beam driven flares using \radyn, and then generated the synthetic \ion{Si}{iv} emission in two ways. The first was the standard optically thin synthesis using CHIANTI contribution functions and ionisation fractions, assuming equilibrium. The second was to use the minority species version of \radyn, \msradyn, to synthesise the two \ion{Si}{iv} lines including the effects of photoionisation and photoexcitation, non-equilibrium ionisation, opacity, and charge exchange. \msradyn\ takes as input certain hydrodynamic variables at at the cadence of \radyn's internal timestep (i.e. the relevant timescales to capture dynamics, not simply the output cadence), and then solves just the NLTE non-equilibrium radiation transfer for a given minority species. That is, there is no feedback of the radiation from the minority species on the atmosphere itself. The line profiles from each method were quite different, even from the pre-flare where charge exchange broadened the \ion{Si}{iv} ion fraction stratification, which peaked slightly cooler in temperature than it does in ionisation equilibrium ($T\sim66$~kK versus $T\sim80$~kK). Charge exchange is generally not considered in transition region line modelling, but both the results of \cite{2019ApJ...871...23K}, and recent quiet Sun modelling of \cite{2021MNRAS.503.1976D,2021MNRAS.505.3968D} highlight its importance. Weaker simulated flares generally produced similar results from both synthesis methods. However, in stronger flares they differed. The peak intensities of \msradyn\ profiles were smaller, but the lines broader overall due to opacity effects so that the integrated line intensities were higher. Those profiles also showed stronger asymmetries, and self-absorption features. Crucially, the intensity ratio deviated from the optically thin ratio of $R_{1394/1402} = 2$. Opacity effects were present in all simulations with an energy injection $F>5\times10^{10}$~erg~s$^{-1}$~cm$^{-2}$, and for some weaker flares with softer electron distributions since they more easily heated the upper chromosphere and lower transition region. Some of these flares only exhibited opacity effects for a transient period, since the transition region and upper chromosphere compressed quickly, meaning there was not a sufficient column mass of \ion{Si}{iv} to build up opacity. When there was an extended flaring lower-transition region (that is temperatures climbing through $30$~kK$<T<100$~kK over a large height range before sharply rising to MK coronal temperatures), opacity effects were present. Roughly, opacity effects were present when the temperatures were enhanced to $40<T<100$~kK above a column mass $5\times10^{-6}$~g~cm$^{-2}$.

There have since been a number of observations of $R_{1394/1402}$ deviating from the optically thin limit $R_{1394/1402} = 2$ \citep[e.g.][]{2021MNRAS.504.2842M,2022ApJ...926..223Z}. \cite{2021MNRAS.504.2842M} found $R_{1394/1402} \ne 2$ at several locations along flare ribbons in an M7.3 flare. \cite{2022ApJ...926..223Z} report similar results, noting also that the ratio varies across the line profile, with stronger opacity in the core so that photons scattered from an optically thick core can easily escape through optically thin line wings. In those observations, we might infer that the flaring atmosphere produced the extended region of $40<T<100$~kK at sufficiently high density. It is important to note that if we do not see much observational evidence for these potentially short lived opacity effects, then our models may be predicting too much density at these intermediate temperatures. Further RT modelling, particularly of other transition region lines in conjunction with \ion{Si}{iv} is sorely required, as are high cadence observations to catch potentially transient opacity effects. Another impact of potential opacity effects in transition region lines, and motivation for their further study, is that these are important contributors to the (assumed) optically thin radiative loss functions, which are a major component energy loss in the simulations, governing the plasma response. 

\cite{2021ApJ...912..121P} and \cite{2021ApJ...915...77P} explored, using machine learning techniques based around mutual information theory\citep[MI;][]{1990JSP....60..823L}\footnote{MI captures statistical dependencies, in this case between various features of different spectral lines. If the information in one spectral lines is independent from the other spectral line, the joint probability of obtaining a certain property of line X alongside a property of line Y is the product of the individual probabilities. If however the probabilities are related due to some correlation between the properties of lines X and Y then the joint probability is a more complicated evaluation. The specific elements of “information” are many and varied (for example the probability that a \ion{Mg}{ii} line has a single peak or central reversal at the same time that \ion{Si}{iv} is doppler shifted). As applied by \cite{2021ApJ...912..121P} and \cite{2021ApJ...915...77P} MI aggregates the many properties and outputs a single score that describes how correlated the lines appear to be overall, which can be studied spatially and temporally.}, the correlation between the various spectral lines of IRIS through the transition region and chromosphere. I encourage the reader to read their detailed analysis carefully, in particular the subtleties surrounding selecting flaring areas and how this might impact correlations, but note the headline results here. They find weak correlations between spectral line pairs during quiescent periods, but substantially enhanced correlations of those pairs during solar flares. \ion{Mg}{ii} and \ion{C}{ii} have the strongest correlation, followed by their correlations with \ion{Si}{iv}. Other lines (e.g. \ion{O}{iv}) are more weakly correlated, and others such as \ion{Fe}{ii} only show strong correlations directly over flare ribbons. This coupling meant that \cite{2021ApJ...915...77P} were able to predict the most probable spectrum of a certain IRIS observable given an input \ion{Mg}{ii} spectra, for example. The strong correlation of \ion{Si}{iv} to the chromospheric lines, despite the weak correlation of other transition region lines such as \ion{O}{iv}, could be due to the deeper formation height suggested during the \cite{2019ApJ...871...23K} simulations. Further, the coherency that flares introduce could be a result of the strong compression of the chromosphere and transition that occurs in many flare simulations. For example, Figure 11 in \cite{2019ApJ...871...23K} shows that over time the range of formation height of the IRIS line cores can shrink to a very small $\Delta z$. The `big data' studies of Dr. Panos and collaborators provide an excellent test bed against which models can be critiqued -- our models should be able to produce similar coherency between the various lines observed by IRIS, and this should be a target of our efforts in the near future.

Finally, I note briefly that it is typical in flare simulations from \hydrad, \radyn, and \flarix\ to produce large enhancements in electron density through the chromosphere and into the corona. These can be in excess of $n_{e} > 10^{13-14}$~cm$^{-3}$ in the chromosphere, and  $n_{e} > 10^{10-12}$~cm$^{-3}$ through the transition region and lower corona. Indeed, as discussed in the preceding sections, a very large electron density at the \ion{Mg}{ii} formation temperatures is required to explain the single peaked profiles. IRIS and \textsl{Hinode}/EIS density sensitive lines from the corona and transition region can demonstrate if these densities are consistent with observations. \cite{2016ApJ...816...89P} measured the ratio of the \ion{O}{iv} 1399.77~\AA\ and \ion{O}{iv} 1401.16~\AA\ line pair, which form at $T\sim158$~kK, during the impulsive phase of the X2 class flare that occurred on 2014-October-27th. The ratio reached the high-density limit, indicating that the density of the flare transition region reached $n_{e} > 10^{12}$ cm$^{-3}$. A caveat here is the assumption of ionisation equilibrium, so that the observed ratio may be in part due to non-equilibrium effects. Other assumptions are that the lines are free of unknown blends, and that the plasma is a Maxwellian, which may not be the case in solar flares, or even active regions, which have been seen to exhibit $\kappa$ distributions \citep[e,g,][]{2016A&A...590A..99J,2018ApJ...853..158D,2022arXiv220707026D}. Similar analysis using EUV spectral lines from EIS indicated a coronal density at $2$~MK of $n_{e} > 10^{10-11}$~cm$^{-3}$. \cite{2016ApJ...816...89P} then modelled this flare using \hydrad, finding that the electron density in the synthetic flaring atmosphere (both flare footpoints and the transition region/lower corona) were consistent with the observationally derived values. 

Pivoting slightly to white light observations, the IRIS NUV Balmer continuum modelling and observations seem to suggest that there is likely some contribution to the optical continuum excess in flares from recombination radiation in the upper chromosphere (see Section~\ref{sec:continuum}). Given the dependence of bound-free (and also free-free) emission on electron density, NUV and optical continuum observations of the chromosphere can give us some means to investigate the density there, and off-limb observations allow us to isolate the chromospheric portion. Off-limb observations of white light flares have revealed both the typical footpoint sources at the base of flare loops as well as bright loop structures (also referred to as prominence loop systems in some literature). The former was discussed by \cite{2017ApJ...847...48H}, who analysed SDO/HMI continuum data of off-limb flares that revealed co-spatial HMI 6173~\AA\ and RHESSI hard X-ray emission, with a characteristic height of $\sim1000$~km \citep[see also][]{2015ApJ...802...19K}. Using an analytical argument of the relative strength of white-light continuum emission mechanisms, \cite{2017ApJ...847...48H} determined that for electron densities above $n_{e} > 10^{12}$~cm$^{-3}$ Balmer bound-free recombination emission dominated over Thomson scattering of incident radiation from the solar disk, with some contribution from free-free emission. Further comparisons using \flarix\ electron-beam driven flares confirmed their supposition, with the simulations containing a high electron density ($n_{e} > 10^{12-13}$~cm$^{-3}$) at the height range of the observed HMI emission, and with an intensity as a function of height resembling the HMI observations (with some assumed loop thickness). These results are consistent with the IRIS NUV continuum flare footpoint observations. To my knowledge no off-limb IRIS NUV flare observations have been reported, and it is likely they would be fairly weak unless a long exposure time was used, but those would be very interesting to compare to the HMI sources. Looptop structures that are readily apparent in optical continuum observations pose a bit more of a challenge for models to reproduce, namely due to the very high coronal densities they imply. Several studies of very strong flares have inferred looptop electron densities between $n_{e} = 10^{12} - 10^{13}$~cm$^{-3}$, usually in the gradual phase of flares, presumably once loops have cooled. For example, \cite{1992PASJ...44...55H} studied both footpoint and loop sources in the the 16th August 1989 flare that was estimated to be an X20 class event. They predicted the intensity of emission from Thomson scattering, free-free, or recombination radiation for a range of temperatures given an assumed emitting volume, inferring from the observed intensity that $n_{e} = [5\times10^{11}, 2\times10^{12},1\times10^{13}$~cm$^{-3}$ for $T = [10^{4},10^{5},10^{6}]$~K, respectively, in a loop source several hundred km above the white-light footpoint sources. A similar analysis was performed by \cite{2018ApJ...867..134J} using HMI observations from the X8.2 10th September 2017 event, finding $n_{e} = 10^{12} - 10^{13}$~cm$^{-3}$ were the most likely values in a large parameter space of temperatures and emitting thicknesses. Inverting \ion{Ca}{ii} 8542~\AA\ and H~$\beta$ data taken by the Swedish Solar Telescope of that same flare, \cite{2019ApJ...885..154K} found consistent values in the cool loops. By studying polarisation of HMI data from the X2.8 flare on 13th May 2013, \cite{2014ApJ...786L..19S} determined that the emission could not be solely due to Thomson scattering and estimated an electron density in the range $n_{e} = 3.5\times10^{11} - 1.8\times10^{12}$~cm$^{-3}$. While flare models can readily explain electron densities up to a few $\times10^{11}$~cm$^{-3}$ in the upper portion of the corona, obtaining higher electron densities at looptops is less straightforward and demands an explanation.

%%%%%%%%%%%%%%%%%%%%%%%%%%%%%%%%%%%%%%%%%%%%%
%% ENERGY TRANSPORT BEYOND PRE-IRIS
%%%%%%%%%%%%%%%%%%%%%%%%%%%%%%%%%%%%%%%%%%%%%

\section{Energy Transport in Flares}\label{sec:mechs}
In this section I discuss how IRIS observations are aiding our efforts to not only refine and challenge the details of the electron beam model, but also in our efforts to explore additional energy transport mechanisms. Alternative mechanisms, that may act in concert with, or instead of, non-thermal electrons (likely varying in dominance in different spatial locations) that are under active study are: non-thermal protons or ions, downward propagating Alfv\'enic waves, and conductive heat flux resulting from direct \textsl{in-situ} heating of the corona. There are possibly others too! I do not touch on non-thermal protons or heavier ions here, other than to say it that these accelerated ions are undoubtedly produced during solar flares and that they may carry energy equivalent to that of electrons \citep[][]{2000AIPC..522..401R,2009ApJ...698L.152S,2012ApJ...759...71E,2017ApJ...836...17A}. That means we could be missing up to half of the energy delivered to the lower atmosphere in flares! \cite{2020ApJ...902...16A} recently updated the \fp\ code, which has been merged with \radyn, to model the propagation of these suprathermal ions, and initial results have demonstrated that protons can penetrate much deeper into the lower atmosphere than electrons, aided by warm target effects \cite[e.g.][]{2021AGUFMSH24B..04A}, and I look forward to studies that use \radynfp\ proton-beam driven flares to forward model IRIS observables. 

\subsection{Coronally-Generated Alfv\'enic Waves in Flares}
First proposed as a means of heating the temperature minimum region where non-thermal electrons likely could not reach, but which observational evidence suggested experienced a modest temperature rise in flares, \cite{1982SoPh...80...99E} constructed a simple but informative model of energy transport via downward propagating, coronally-generated, Alfv\'enic waves. In this model, waves would be produced from the reconnection site, propagating through the corona into the lower atmosphere to the temperature minimum region where they were damped by resistivity. These simulations assumed mono-chromatic (single frequency) waves, employed the WKB approximation (that is, waves were not reflected by density gradients), and assumed an instantaneous travel time. These assumptions allowed a straightforward formulation of a damping length to model the dissipation. 

This notion was revisited by \cite{2008ApJ...675.1645F} who investigated the possibility that Alfv\'enic waves could not only deliver the energy liberated by magnetic reconnection to the chromosphere, and accelerate electrons in the corona via field-aligned electric fields but could also potentially locally accelerate electrons in the chromosphere via mode-conversion to high wave-numbers resulting in turbulent acceleration. More work is required to understand the role of these waves in particle acceleration. The thought experiments of \cite{2008ApJ...675.1645F} explored Alfv\'enic waves as an alternative to the electron beam model as a means to deliver flare energy and explain observations of both hard X-rays and broadband enhancements of the UV/optical/infrared. This was motivated by perceived issues with the coronal acceleration problem, namely the vast numbers of electrons required ($>10^{36}$ elec s$^{-1}$), which can quickly deplete the coronal volume of ambient electrons. Return currents can resupply the corona with electrons, however, mitigating this problem.

Even if they are not required as a complete replacement to electron beams (which is still a source of vigorous debate), it is important that we continue to properly consider the role of Alfv\'enic waves in flares. Flares are, fundamentally, a violent restructuring of the magnetic field, meaning that MHD waves are undoubtedly produced. The question is, do they carry sufficient energy to play a non-negligible role in transporting energy compared to coronally accelerated electrons, and can they efficiently heat the chromosphere (either alongside or instead of those electrons). Additionally, we do not see hard X-rays all along the flare ribbons. Perhaps different parts of ribbons are heated by different mechanisms. Some MHD simulations by \cite{2013ApJ...765...81R} and \cite{2013A&A...558A..76R} revealed that Alfv\'enic waves could penetrate the transition region density boundary if they had a high enough frequency, $f> ~1$~Hz, meaning the WKB approximation could be used within loop models to further investigate high-frequency Alfv\'enic waves. They also noted that ion-neutral interactions were important, alongside electron resistivity, in damping the Alfv\'en waves. 

Inspired by these results \cite{2016ApJ...818L..20R} modified \hydrad\ to model Alfv\'en waves using the WKB approach of \cite{1982SoPh...80...99E}, but with an updated treatment of damping which included ambipolar effects. Thus, the waves were damped by ion-neutral, neutral-electron, and electron-ion collisions. Modelling a range of Alfv\'en wave parameters, including the injected Poynting flux, mono-chromatic frequency, and wave number they found that they could strongly heat the chromosphere, and that they could drive explosive chromospheric evaporation. This model was further improved in \hydrad\ by \cite{2018ApJ...853..101R}, to include the wave travel time, via ray-tracing so that the waves propagate at the local Alfv\'en speed. They show that in addition to certain wave parameters being damped more effectively in the lower atmosphere than in the upper chromosphere, that leading waves can effectively bore a hole through the chromosphere allowing following rays to penetrate deeper into the lower atmosphere. This occurred due to ionisation by the leading waves, reducing the local damping. 

Following the approach of \cite{2016ApJ...818L..20R}, \cite{2016ApJ...827..101K} included Alfv\'en waves as a mode of energy transport into \radyn. This initial work employed the instant-travel approximation where the wave propagation was ignored. I have since updated \radyn\ to include the travel time of the wave in the same manner as \cite{2018ApJ...853..101R}. For the remainder of this section I mostly discuss the results using the \cite{2016ApJ...827..101K} model, since those are the published results relevant to IRIS, but work modelling the IRIS observables including travel time is underway via both \hydrad\ and \radyn. 

\cite{2016ApJ...827..101K} compared the atmospheric dynamics and radiative output of two \radyn\ simulations, (1) an electron beam, and (2) a mono-chromatic Alfv\'en wave. The energy flux of each was set to be $1\times10^{11}$~erg~s$^{-1}$~cm$^{-2}$, and the Alfv\'en wave parameters set to most effectively heat the upper chromosphere. A magnetic field stratification was imposed for the purpose of defining the Alfv\'en speed and damping lengths; it did not evolve during the simulations. Two spectral lines were compared, the \ion{Ca}{ii} 8542~\AA\ and \ion{Mg}{ii} k line, the latter synthesised using \radyn\ atmospheres with \rh. While the atmospheres showed some striking similarities in each model's ability to heat the chromosphere and drive strong upflows, as was first seen in \cite{2016ApJ...818L..20R}, there were intriguing differences in the chromospheric stratification. These differences revealed themselves in the spectral lines also. 

The Alfv\'en waves produced a flatter, more spatially extended energy deposition profile, resulting in temperature rises at deeper heights than the electron beam simulation. Despite this, it did take time for the electron density in the lower atmosphere to catch up to the electron beam simulation because of the absence of non-thermal collisional ionisation due to the beam itself. Runaway helium ionisation due to the more concentrated electron beam heating removed the $304$~\AA\ line as a radiator, resulting in a high temperature bubble forming, flanked by narrow cool high density regions. These flanking regions expanded as a high velocity upflow, and slower downflow (in addition to the initial explosive evaporation). A secondary upflow appeared in the Alfv\'en wave simulation also, but was more gentle with a shallower spatial gradient. In the electron beam simulation the \ion{Mg}{ii} k line had a central reversal that was redshifted during most of the heating phase. The line wings had small optically thin contributions due to the flow patterns. In the Alfv\'en wave simulation, however, the line formed in a gentle upflowing region of the chromosphere, shifting the absorption profile strongly to the blue. Since the densities in the upflow were relatively weak this did not fully shift the line but instead pushed the core and blue k2v peak closer in formation height until they merged. The upflow produced optically thin contributions through the blue wing. Fewer absorptions due to shifting the absorption profile boosted the red k2r peak in comparison to the heavily suppressed k2v peak, meaning that the whole profile took on a very asymmetric form. The k line could be mistaken as being single peaked with a large blue wing asymmetry. Differences in the shape of the \ion{Mg}{ii} k line cores were a direct result of the different flows, that themselves were due to the different stratifications of damping in either the electron beam or Alfv\'en wave energy transport mechanisms. \cite{2016ApJ...827..101K} demonstrated that \ion{Mg}{ii} can help discriminate between energy transport models, but much more work needs to be done here, particularly studying multiple IRIS spectral lines forming in a wider array of Alfv\'en wave driven simulations that include the wave travel time. The predictions from each model should also be compared to the $k$-means classifications of \cite{2018ApJ...861...62P}, and we must work to improve the models to include a spectrum of wave frequencies, and to constrain the properties of the waves.

\begin{figure}[h]
\begin{center}
\includegraphics[width=\textwidth]{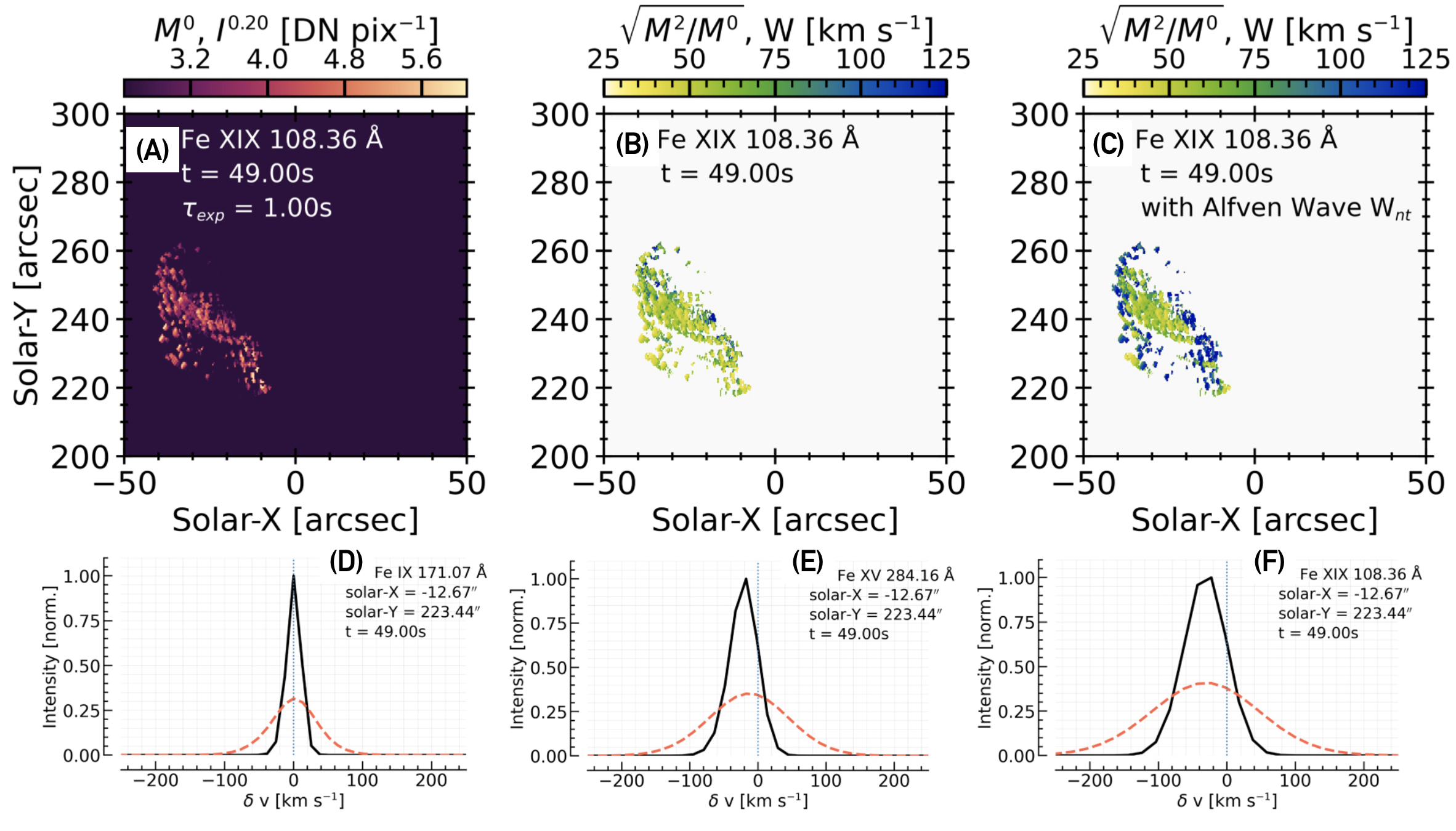}% This is a *.eps file
\end{center}
\caption{Demonstrating how Alfv\'en waves may explain some of the anomalous broadening of hot flare lines. In this \radynarcade\ flare simulation predictions were made of the MUSE 108~\AA\ line, forming at $10$~MK. Panel \textbf{(A)} shows the a map of the line intensity (zeroth spectral moment), scaled to the 1/5th power to show both weak and strong sources, at $t=49$~s into the simulation, where hot footpoints and loop legs are apparent. Panel \textbf{(B)} shows a map of the line widths (second spectral moment) where broadening is solely due to thermal and instrumental effects and the superposition of sources along the line of sight. Panel \textbf{(C)} also shows a map of line width, but also includes broadening due to an Alfv\'en wave propagating along each loop, with a Poynting flux of $1\times10^{10}$~erg~s$^{-1}$~cm$^{-2}$ (a magnetic field was assumed for the purposes of calculating the Alfv\'en speed). Clearly the line was much broader. Panels \textbf{(D-F)} show individual spectra, where black is the original, and the red-dashed is the Alfv\'en wave broadened version. Figure adapted from \cite{2022ApJ...926...53C}. \copyright AAS. Reproduced with permission. \copyright AAS. Reproduced with permission.} \label{fig:alfven}
\end{figure}

While Alfv\'en waves are certainly produced during magnetic reconnection \citep[they have detected \textsl{in-situ} in the magnetosphere, e.g.][]{2005PhRvL..95f5002C,2002JGRA..107.1201W,2017NatCo...814719G}, a vital question is how much energy do they carry to the lower atmosphere? Is it enough to compete with electron beams as an important contributor to the flare energy budget, or is it negligible and thus safely ignorable? Thus far, the simulations of \cite{2016ApJ...818L..20R,2018ApJ...853..101R} and \cite{2016ApJ...827..101K} injected a Poynting flux of the level that we know from electron beam driven flare simulations, and bolometric flare observations, is required to significantly heat the chromosphere. An observational constraint on the Poynting flux is required. An upper limit could be placed on this by investigating the width of lines formed at different temperatures (i.e. altitudes). The non-thermal component of the width could result from ion motion in response to an Alfv\'en wave. To demonstrate what the upcoming EUV observations Multi-Slit Solar Explorer \citep[MUSE, scheduled for launch in $\sim$2026;][]{2020ApJ...888....3D} would reveal about solar eruptive events, many flare models synthesised MUSE observables and demonstrated how MUSE might discriminate between model predictions \citep{2022ApJ...926...53C}. As part of that effort we modelled the broadening that would be induced due to an Alfv\'en wave propagating down the loops in our \radynarcade\ model, noting that the line was indeed substantially broadened. This is demonstrated in Figure~\ref{fig:alfven} which shows the \radynarcade\ model before and after Alfv\'en wave broadening is included. Coordinated high spatiotemporal resolution observations between MUSE and the High Throughput EUV Solar Telescope (EUVST, also scheduled for a $\sim$2026 launch) could track the development of non-thermal widths during a flare, placing constraints on the Poynting flux. Knowledge of the coronal magnetic field would also be very advantageous here, to help set the Alfv\'en speed and damping lengths, and to determine the amplitude of magnetic field perturbations. Finally, it is worth noting that Alfv\'enic waves have been proposed as the mechanism responsible for the observed elemental fractionation between the photosphere and corona. Low-first ionisation potential (FIP; $<10$~eV) species generally have a coronal abundance that is 4 or so times that of the photosphere. The ponderomotive force generated in MHD waves has been suggested as a potential cause of the so-called FIP effect \citep[e.g.][]{2015LRSP...12....2L}. Observations of the FIP effect in flares \citep[e.g.][]{2018ApJ...853..178D} may then shed light on the properties of Alfv\'enic waves produced during flare reconnection.
 
\subsection{High Non-Thermal Electron Energy Fluxes}
The energy flux injected to dynamic flare simulations has typically ranged on the order $F = 10^{9-11}$~erg~s$^{-1}$~cm$^{-2}$, driven in part, admittedly, because of the computational expense and difficulty of injecting very much stronger values of $F$ into time-dependent models until fairly recently ($F>10^{12}$~erg~s$^{-1}$~cm$^{-2}$ fluxes, while computationally demanding, are now possible). This range has been inferred from numerous studies of flares in both the RHESSI era and before, but we are now realising that in some of the strongest flare sources we may be underestimating $F$, perhaps by an order of magnitude in some cases! The physical rationale and implications behind this, with regard to non-thermal particle production and transport, are beyond the scope of this review, but an important factor in tying down the existence of very high beam fluxes are the modern observations at high spatiotemporal resolution  of UV and optical flare sources. IRIS, \textsl{Hinode}/Solar Optical Telescope (SOT), and ground based observatories have revealed flare sources are smaller that typically assumed from older data (particularly so if looking at white light flare data)\footnote{An assumption here is that the white light flare area represents the same area into which electrons are deposited. While the hard X-ray sources sizes are large due to the relatively poor spatial resolution of those instruments, we do not know with certainty that the small white light areas represent the true areas from which hard X-rays originate. There is some ambiguity as to the relative heights at which hard X-rays and white light emission is produced, though as we will see models do suggest they are close. Observations of limb flares also suggest that some white light emission and the hard X-rays come from the same volume \citep[e.g.][]{2015ApJ...802...19K}. Additionally, there could be area expansion through the loop affecting source areas. All this is to say that while we now believe that white light and UV source areas are truer representations of the area into which non-thermal electrons are deposited, it is perhaps best to say that they are a lower limit, with the upper limit coming from the hard X-rays.}. A detailed study comparing source sizes from flare ribbons observed by \textsl{Hinode}/SOT to hard X-ray imaging spectroscopy suggested that the beam flux may very well be $F>10^{12}$~erg~s$^{-1}$~cm$^{-2}$ in that flare \citep{2011ApJ...739...96K}. Further, some groups have started looking at newly activated sources to define the areas into which energy is being injected within some observational window. Newly activated sources might be as small as to be on the order $10^{16}$~cm$^{-2}$ or below \citep[][]{2011ApJ...739...96K,2014ApJ...788L..18S,2014ApJ...793...70M,2016ApJ...816...88K,2017ApJ...836...12K,2020ApJ...895....6G}. IRIS observations can both guide the magnitude to inject based on high resolution observations of source areas, and act as a validation. 

\cite{2017ApJ...836...12K} injected fluxes of $F=[1\times10^{11},5\times10^{11}]$~erg~s$^{-1}$~cm$^{-2}$ to simulate the two brightest sources in the 2014-March-29th X class flare, focussing on modelling the NUV continuum response. These fluxes were guided by the hard X-ray analyses of \cite{2016ApJ...816...88K} and \cite{2015ApJ...813..113B}, with the range based on arguments of the continuum emitting areas identified by \cite{2017ApJ...836...12K}. Portions of the NUV continuum in the region $\lambda \sim [2814-2832]$, observed by IRIS, were first identified by \cite{2014ApJ...794L..23H}, who extracted patches of continua free from lines. They determined these line-free regions as likely being part of the Balmer continuum that remained optically thin during the flare and which formed in the mid-upper chromosphere. This means the continuum response would be very sensitive to the electron density throughout the flare chromosphere.  \cite{2017ApJ...836...12K}'s numerical experiments showed that the NUV continuum was too weak in the lower energy flux simulation, and \textsl{much} too weak in a set of experiments in which similar energy flux was instead deposited directly in the corona and allowed to conduct down to the chromosphere (potentially due to the lack of non-thermal collisional ionisations in the conduction-only simulations, though this was not commented on by the authors). In the high energy flux simulation the continuum did reach a sufficient level to match observations by $t\sim2$~s, peaked a few seconds later, before declining thereafter (but still remaining $100-200~\%$ above the pre-flare). Thus, a high energy flux was in fact required to produce conditions to raise the continuum intensity to the observed level. An in-depth analysis found that the NUV continuum was formed by hydrogen recombination emission from two distinct layers, both optically thin: a stationary chromospheric layer and a dense condensation that rapidly forms and accrues mass. As time progressed the condensation became responsible for the bulk of the emission, due to the fact that as the density increased, an increasing proportion of the non-thermal electrons thermalised in the condensation itself, and consequently the stationary layer cooled somewhat. The conditions inside this condensation were found to be comparable to those of earlier slab model explanations of Balmer continuum enhancements \citep[e.g.][]{1985A&A...152..165D}, suggesting that condensations (and high beam fluxes) are required to explain the brightest continua enhancements. 

Various effects should be accounted for when considering very large non-thermal electron flux densities, such as the beam-neutralising return current \citep[including the effects of runaways;][]{2005A&A...432.1033Z,2012ApJ...745...52H,2020ApJ...902...16A,2017ApJ...851...78A,2021ApJ...917...74A}, and instabilities that affect the beam propagation \citep[e.g.][]{2009ApJ...707L..45H,2008A&A...478..889L,2014ApJ...793....7L}. A discussion of those is included in \cite{2017ApJ...836...12K}, but are beyond the scope of this review, though I note that the careful model-data analysis of the type performed by \cite{2017ApJ...836...12K} is crucial as we explore the impact of these effect in flare loop models. 

\subsection{Constraining Flare Energetics with Balmer Continuum Observations}\label{sec:continuum}
In the standard flare model there is typically not enough power carried by the highest energy electrons to meaningfully heat the deepest chromospheric/photospheric layers. However, there are some strands of evidence that suggest we do indeed require heating deeper than models currently predict. Listing some examples: excess line widths of chromospheric transitions (e.g. \ion{Mg}{ii} h \& k) cannot be accounted for; there is evidence of heating at the temperature minimum region (e.g. from inversions of \ion{Mg}{i} $\lambda\lambda4571$\AA\ \& $\lambda\lambda5173$\AA, \citealt{1990ApJ...350..463M,1990ApJ...365..391M}); there is speculation that white light flares (WLFs) may originate from the photosphere via enhanced H$^{-}$ emission following a local temperature increase, or contain significant contributions from the lower atmosphere \citep[see discussions in][]{1989SoPh..121..261N,1989SoPh..124..303M,1993SoPh..144..169N,2014ApJ...780L..28M,2014ApJ...783...98K,2016ApJ...816...88K,2018A&A...620A.183J}. An alternative explanation to the WLF problem is optically thin bound-free (recombination) radiation resulting from overionisation of the mid-upper chromosphere \citep[e.g.][]{1972SoPh...24..414H}. This would produce a Balmer jump at $3646$~\AA. Of course, likely both of those mechanisms play a role. If we do need heating to great depth, then we must identify the agent capable of doing so, and constrain how much energy is required.

Given the scarcity of white light continuum observations, the NUV continuum as observed by IRIS is one such means to constrain the need for deep heating (the NUV is thought to be closely related to the optical continuum, albeit we do not yet know if they always originate from the same volume during flares). \cite{2014ApJ...794L..23H} first determined that the Balmer continuum could be observed by IRIS, using observing windows in the NUV near $\lambda = [2813-2816,2825-2828, 2831-2834]$~\AA. They carefully extracted narrow, line free, portions of the spectrum, finding $\sim100-200$~\% contrast compared to the pre-flare values, with an impulsive rise and more gradual decay. Comparing to bright, non-flaring features, they note that the continuum rose but lines did not (as in the flare case) suggesting that the continuum patches between the lines are unaffected by the line emission. Using the static flare models of \cite{1983ApJ...272..739R}, processed with the radiation transfer code \mali, they inferred from the similarity in model-to-data intensities and from the formation properties in the model, that the observed NUV spectra did represent the Balmer continuum, and that it was due to optically thin recombination emission in the upper chromosphere. This represented the first detection of the Balmer continuum from space based instruments, and provides a constraint on flare models due its proximity to the Balmer jump. 

\begin{figure}[h]
\begin{center}
\includegraphics[width=\textwidth]{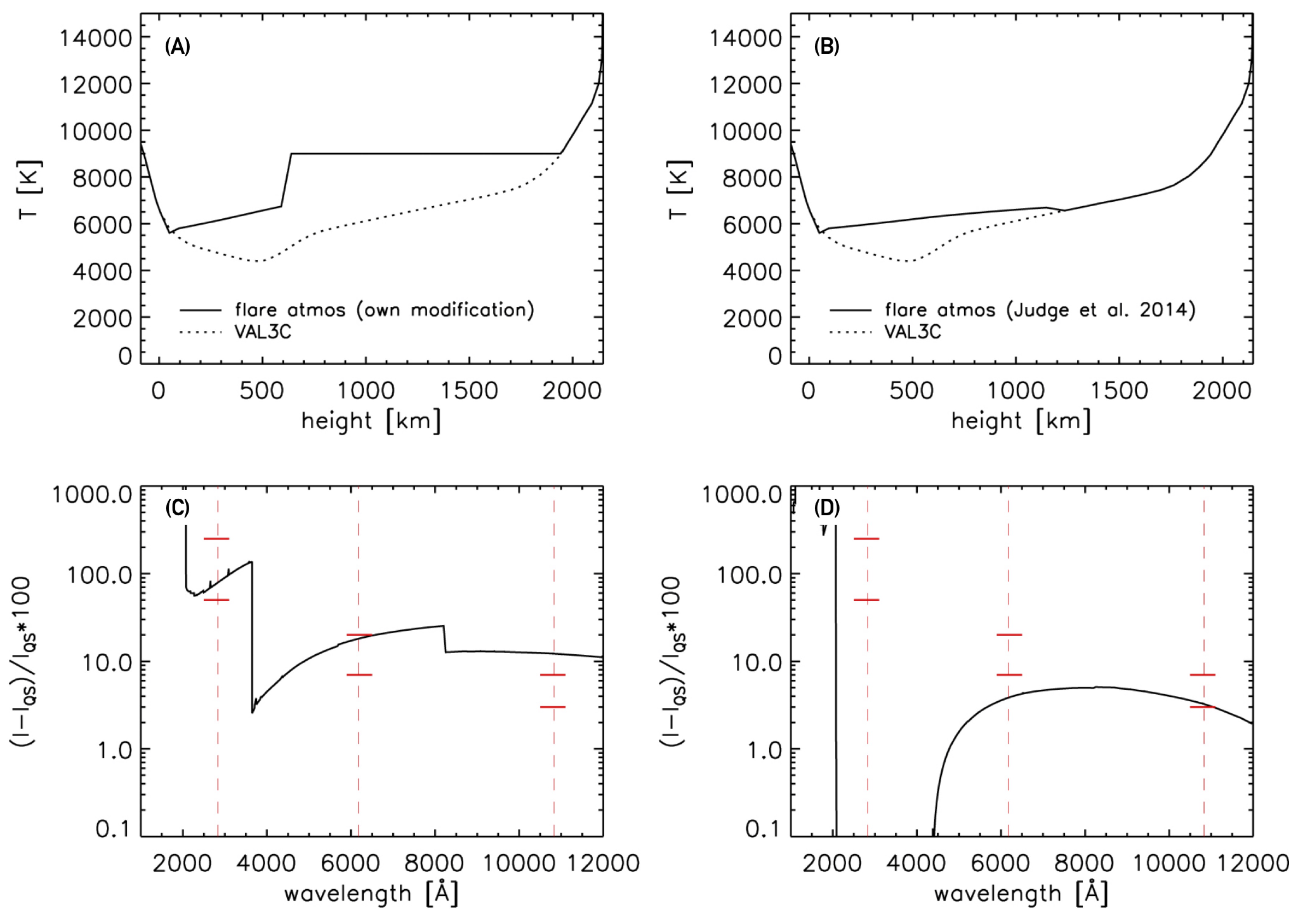}% This is a *.eps file
\end{center}
\caption{Synthetic continuum spectra from flare atmospheres processed using \rh. Here \cite{2016ApJ...816...88K} manually modified the temperature in the lower atmosphere to demonstrate that both a chromospheric temperature enhancement, and a temperature enhancement at greater depth (panel \textbf{A}) was required to synthesise the NUV and optical continuum (black line in panel \textbf{C}) that was consistent with the IRIS and SDO/HMI observations (red symbols on panel \textbf{C}). A lower atmosphere increase on its own (panel \textbf{B}) was not able to explain the Balmer continuum observations from IRIS (panel \textbf{D}). Figure adapted from \cite{2016ApJ...816...88K}. \copyright AAS. Reproduced with permission.} \label{fig:continuum}
\end{figure}

Following on from this initial detection, \cite{2016ApJ...816...88K} studied the IRIS Balmer continuum emission alongside other continuum enhancements from both space and ground observatories, spanning the UV through infrared. They performed blackbody fits to data (including modifying the blackbody intensity due to opacity effects) to determine the viability of an upper photospheric origin to the continuum emission, finding that the NUV lay well above the blackbody curve predicted by the optical and IR emission (as expected if the NUV emission was indeed recombination radiation, producing a Balmer jump). Building upon the modelling work started by \cite{2014ApJ...794L..23H}, \cite{2016ApJ...816...88K} selected a few models from \cite{1983ApJ...272..739R} and calculated the non-LTE hydrogen recombination continuum using the \mali\ code. Several were consistent with the observed NUV continuum enhancements (with a non-thermal electron flux close to that derived from RHESSI observations for that flare), but those models under-predicted the optical and IR enhancements from SDO/HMI and the Facility Infrared Spectropolarimeter at the Dunn Solar Telescope (DST/FIRS).  Instead, a semi-empirical model atmosphere with photospheric temperature rise was required to achieve consistency with the optical / IR observations. Finally, they manually modified atmospheres that were input to \rh\ in an attempt to find a stratification consistent with all three regimes (UV, optical and IR). A model with a modest photospheric temperature increase alongside a strongly heated chromosphere was required, as shown in Figure~\ref{fig:continuum}. Thus, IRIS in combination with ground based observations demonstrated that in some events we may indeed require both chromospheric and photospheric heating. To my knowledge no time-dependent flare model has self-consistently produced such an atmosphere (electrons beams typically do not carry enough power to such depths), and this should be a focus of our continuing efforts. However, it should also be noted that since an optically thin source at $T\sim10$~kK produces optical emission with a radiation temperature of $4-6$~kK \citep[see e.g.][]{2018ApJ...852...61K}, some ambiguity remains. At the same time as attempting to model self-consistently the heating throughout the chromosphere and photosphere to determine how to obtain atmospheres similar to the empirical models of \cite{2016ApJ...816...88K}, we should endeavour to obtain IRIS observations in the NUV alongside a broad spectral coverage of the optical (e.g. from DKIST) to determine the spectral shape more accurately, which would help resolve the ambiguity over emission mechanisms. 

Since it will likely remain challenging to obtain observations covering the Balmer jump (and thus a guide as to the formation of the optical continuum), \cite{2019ApJ...878..135K} have begun to search for alternative metrics that can gauge the extent to which the lower atmosphere is strongly heated. Using the fact that \ion{Fe}{ii} lines observed by IRIS form under similar physical conditions as the NUV Balmer continuum \citep[$T\sim8-18$~kK but mostly towards the cooler end, discerned from their earlier modelling work regarding high beam fluxes;][]{2017ApJ...836...12K}, they explored the ratio of wavelength-integrated flare excess \ion{Fe}{ii} $2814.45$~\AA\ intensity to the average continuum intensity in the region $\lambda = [2824.5 - 2825.9]$~\AA. This ratio was observed to be $R_{Fe:NUV}\sim7-8$ at the peak of very bright flare sources located in a sunspot umbra during the 2014-Oct-25th X class flare, significantly higher than the prediction from slab models with low-to-moderate densities of $\rho < 10^{-9}$~[g~cm$^{-3}]$ which had values $R_{Fe:NUV}\sim1$. They speculate that this means there is significant heating (to $T\sim10$~kK) at high column depth ($\mathrm{log}~m \sim -2$~[g~cm$^{-2}$]) where \ion{Fe}{ii} can be optically thick. They are currently modelling this ratio in a range of \radyn\ flares (\textsl{private communication}), but noted that their earlier study of the 2014-March-29th X class flare \citep[][]{2017ApJ...836...12K} only produced an observed ratio of $R_{Fe:NUV}\sim1$, with a modelled ratio of $R_{Fe:NUV}\sim1-1.8$. Clearly there is something quite different at work during the 2014-Oct-25th flare. This could be due to the pre-flare atmospheres, since the 2014-October-25th flare sources propagated into the Sunspot umbra, allowing a colossal $1000$~\% NUV contrast, and an excess intensity $20\times$ that of the 2014-March-29th flare. Hopefully further observations from a variety of flares, and modelling of a variety of energy inputs (including varying the pre-flare atmosphere) will lead to a firm diagnostic of deep heating during IRIS flares.  

The NUV continuum was forward modelled by \cite{2016IAUS..320..233H}, using one of the short-pulse experiments of \cite{2009A&A...499..923K}. In those \flarix\ simulations, a non-thermal energy flux was injected in a trapezoidal form over time, with peak flux $F = 4.5\times10^{10}$~erg~s$^{-1}$~cm$^{-2}$. The whole pulse was very short, only lasting $3$~s. The resulting Balmer continuum intensity was quite small compared to the observations of \cite{2016ApJ...816...88K}. This was attributed to the magnitude of energy flux deposited in the chromosphere. It was an order of magnitude smaller than that of the flare studied by \cite{2016ApJ...816...88K}. Further, the short duration of this relatively moderate injection meant that evaporation was weak and the pressure in the upper chromosphere did not increase to the level inferred from the best-match \cite{1983ApJ...272..739R} analysed by  \cite{2016ApJ...816...88K}. This could point to the need for either longer electron beam dwell times in large flares, or for a train of short pulses. In those models the hydrogen subordinate continua, in particular the Balmer continuum, were the dominant source of radiative losses throughout the chromosphere, overtaking losses from singly ionised metals such as \ion{Ca}{ii} and \ion{Mg}{ii}, underscoring the importance of IRIS Balmer continuum observations. Since the Balmer continuum is seemingly optically thin, the radiative losses integrated through the continuum formation heights are directly related to the emergent intensity. Thus, the observed excess intensities impose strict constraints on flare energetics. 

The Balmer continuum is also a useful constraint for smaller events where heating is, largely, confined to the upper chromosphere. A `mini-flare' event accompanying a jet was studied by \cite{2020A&A...642A.169J} determining that reconnection occurred at the base of the jet. A small Balmer continuum excess was present in very localised sources, from both IRIS SG spectra, and the SJI 2832~\AA\ images \cite{2021A&A...654A..31J}. Comparing to the analysis of \cite{2016ApJ...816...88K}, who had processed the \cite{1983ApJ...272..739R} atmospheric models using \mali\ to obtain predictions for the hydrogen recombination continuum, \cite{2021A&A...654A..31J} found that a few simulations were consistent with their observations. This balanced the continuum intensity, as well as the brightness of the \ion{Mg}{ii} line cores. The most well matched models had non-thermal electron energy fluxes $F = 1\times10^{9-10}$~erg~s$^{-1}$~cm$^{-2}$, with $\delta = 5$ and $E_{c} = 20$~keV (note that the \citealt{1983ApJ...272..739R} did not sample other values of $E_{c}$ and only a few values of $\delta$ for more energetic flares). From the \textsl{Fermi}/GBM \citep[][]{2009ApJ...702..791M} hard X-ray observations, the injected non-thermal electron distribution was calculated as having an energy flux $F = 6.5\times10^{9}$~erg~s$^{-1}$~cm$^{-2}$ for $E>20$~keV, assuming the area into which the electrons were injected was the same as the continuum enhancement source. While there is some uncertainty in the low-energy cutoff, the parameters derived from the hard X-ray observations are consistent with those models that also provide a well-matched Balmer continuum excess intensity.  \cite{2021A&A...654A..31J} estimate that  $82~\%$ of the intensity in the IRIS SJI 2832~\AA\ images is continuum emission, contrasting the result from similar analysis of an X-class flare that found significant line emission \citep{2017ApJ...837..160K}, which makes sense if we consider the weak energy flux involved that was unable to sufficiently excite the many lines within that part of the spectrum. In summary, small-scale reconnection at the base of a jet seemingly was able to accelerate enough electrons to bombard the upper chromosphere, enhancing the Balmer continuum, but was unable to really effect the lower chromosphere.  

\subsection{Flares Driven by Conductive Heat Fluxes}
While there is unambiguous observational evidence for the presence of non-thermal electrons in a great many solar flares, they are not ubiquitously present. This is true both in the global sense, meaning there are some `thermal' flares that do not exhibit strong evidence of hard X-rays or microwaves of non-thermal origin, and the local sense meaning that hard X-rays sources do not appear uniformly along optical/UV flare ribbons (though this latter case may be related to the dynamic range of most hard X-ray observatories that preclude detection of weak sources alongside strong footpoints). In such flares \textsl{in-situ} heating of the corona results in a conductive heat flux that transports energy to the lower atmosphere, generating the strong heating and mass flows \citep[e.g.][and references therein]{1988ApJ...329..456Z,2009A&A...498..891B,2011SSRv..159...19F,2012ApJ...754...54B,2012A&A...540A..24B,2022A&A...657A..51L}. This is often referred to as `direct heating', a somewhat nebulous term that refers generally to any heating of the corona (either looptops or along the legs of the loop) following the release of energy during magnetic reconnection\footnote{This does not include potential heating of the corona by very low-energy non-thermal electrons that are thermalised in the lower corona, which are already captured by the models. In such cases, a conductive heat flux is present due to this coronal heating, but in this section we refer to coronal heating in flares in the absence of non-thermal particles.}, including the retraction of magnetic loops that produce shocks such as those modelled by \preft. It is indeed likely that some form of direct coronal heating acts alongside non-thermal electrons even in flares with clear evidence of particle acceleration, but the dominance of each mechanism varies from flare to flare and with spatial location. For example, recent results using GOES soft X-rays observations suggest that in a number of flares the corona is rapidly heated to $T\sim10-15$~MK before the onset of evaporation \citep[so-called `hot onsets'][]{2021MNRAS.501.1273H}. Several studies have either modelled flares as being purely driven by thermal conduction \citep[e.g.][to name but a few]{1983ApJ...265.1090C,1986SoPh..103...47M,1991A&A...241..618G} or have contrasted predictions between electron beam and conduction driven flares \citep[e.g.][]{2018ApJ...856..178P,2021ApJ...912..153K,2022ApJ...926...53C}.  This latter exercise should be performed more often, as it is likely that both mechanisms act but the direct heating in the corona is often ignored. As shown in \cite{2022ApJ...926...53C}, the inclusion of direct heating can have impacts on the predicted intensities and Doppler motions of coronal and transition region lines. Some of those effects can only be seen at very high spatial and temporal resolution, such as will be afforded by the MUSE mission (see Section~\ref{sec:conc}). 

I summarise in detail here two recent examples of modelling flares driven purely by a conductive heat flux, and what we can learn about the nature of condensations from those simulations and IRIS observations.

The seminal studies of mass flows in flare models by \cite{1985ApJ...289..414F,1985ApJ...289..434F,1985ApJ...289..434F}, and \cite{1989ApJ...346.1019F} revealed the relationship between flare energy input and the development of both upflows and downflows in the chromosphere. Of those, \cite{1989ApJ...346.1019F} concentrated on chromospheric condensations, developing an analytical model that described the timescales and magnitudes of flare-induced downflows. \cite{1989ApJ...346.1019F} built that equation of motion from generalising various properties that occurred in flare loop models.  Notably, \cite{1989ApJ...346.1019F} discovered that the lifetime of the condensations depend only on the chromospheric conditions, not the energy input, and is $\tau_{life}\approx2(H/g)^{1/2}$, where $H$ is the chromospheric density scale height and $g$ is gravitational acceleration. For reasonable values of $H$, $\tau_{life}\sim60$~s. The half-life of the condensation was $\tau_{1/2} \simeq 2(H/g)^{1/2}/M_{peak}$, where $M_{peak}$ is the ratio of the peak downflow velocity to the sound speed in the pre-flare chromosphere. Though $\tau_{life}$ does not depend on properties of the energy injection, the peak downflow velocity (and thus $\tau_{1/2}$) does, varying as  $u_{0} \propto F^{1/3}$, with some dependence on whether energy was transported via non-thermal electrons or was conducted down from a hot corona.

\cite{2021ApJ...912...25A} used the \preft\ gas-dynamic code to study conduction-only driven chromospheric condensations, building upon and complementing the work of \cite{1989ApJ...346.1019F}. They found similar relationships, with some differences as described below, but using an alternative approach. They set up a simplified set of physical parameters to explore the dynamics of shocks in the chromosphere using jump conditions, and compared those predictions to numerical experiments. The analytical model informed a subsequent analysis and interpretation of  numerical results from \preft. Unlike \cite{1989ApJ...346.1019F}, \cite{2021ApJ...912...25A} allowed $H$ to be a free parameter in their model, from which they found a similar scaling for the half-life, but with a different factor (2.8 versus 2). Casting in terms of $u_{0}$ this was: $\tau_{1/2}\approx2.8H/u_{0}$. \preft\ was set up as a rigid flux tube so that the only energy transport was via energy injected to the loop to mimic direct flare heating, and conduction-driven flares were simulated. Using those simulations with various values of pre-flare $H$ and injected energy flux $F$, the velocity as a function of time $u(t)$ was fit with a similar functional form as that from their analytical model. The best-fit values suggested that the relationship was instead $\tau_{1/2}\approx1.67H/u_{0}$, with the discrepancy attributed to at least one of the simplifying assumptions in their analytical model not being satisfied. The results of \cite{1989ApJ...346.1019F} and \cite{2021ApJ...912...25A} demonstrate that both analytically and numerically the pre-flare chromospheric density scale height $H \propto u_{0}\tau_{1/2}$, with the numerical results of \cite{2021ApJ...912...25A} pointing to $H \simeq 0.6u_{0}\tau_{1/2}$. The slight difference between the results of \cite{2021ApJ...912...25A} and the earlier work of \cite{1989ApJ...346.1019F} could be due in part to the heating profile assumed. \cite{1989ApJ...346.1019F} assumed short pulses of non-thermal electrons in the models which the analytical expressions used a base, whereas \cite{2021ApJ...912...25A} assumed direct heating in the corona. Nevertheless, this is a rather powerful diagnostic that suggests variations in the lifetime of condensations could be related to variations in the pre-flare chromospheric densities into which shocks propagate, and that variations along a flare ribbon could reveal corrugation of the pre-flare atmosphere. Note that the condensations referred to here are the relatively strong, transient, downflows that may appear on top of a longer-lived envelope as discussed Paper 1. The peak velocity, $u_{0}$, itself scaled with the input energy flux, without much reliance on $H$, going as $u_{0} \propto F^{1/2}$ for weaker energy fluxes ($F<2\times10^{10}$~erg~s$^{-1}$~cm$^{-2}$), and  $u_{0} \propto F^{1/3}$ for stronger fluxes. The latter was predicted by \cite{1989ApJ...346.1019F}, and the former by \cite{2014ApJ...795...10L} for low-energy fluxes.

\begin{figure}[h]
\begin{center}
\includegraphics[width=\textwidth]{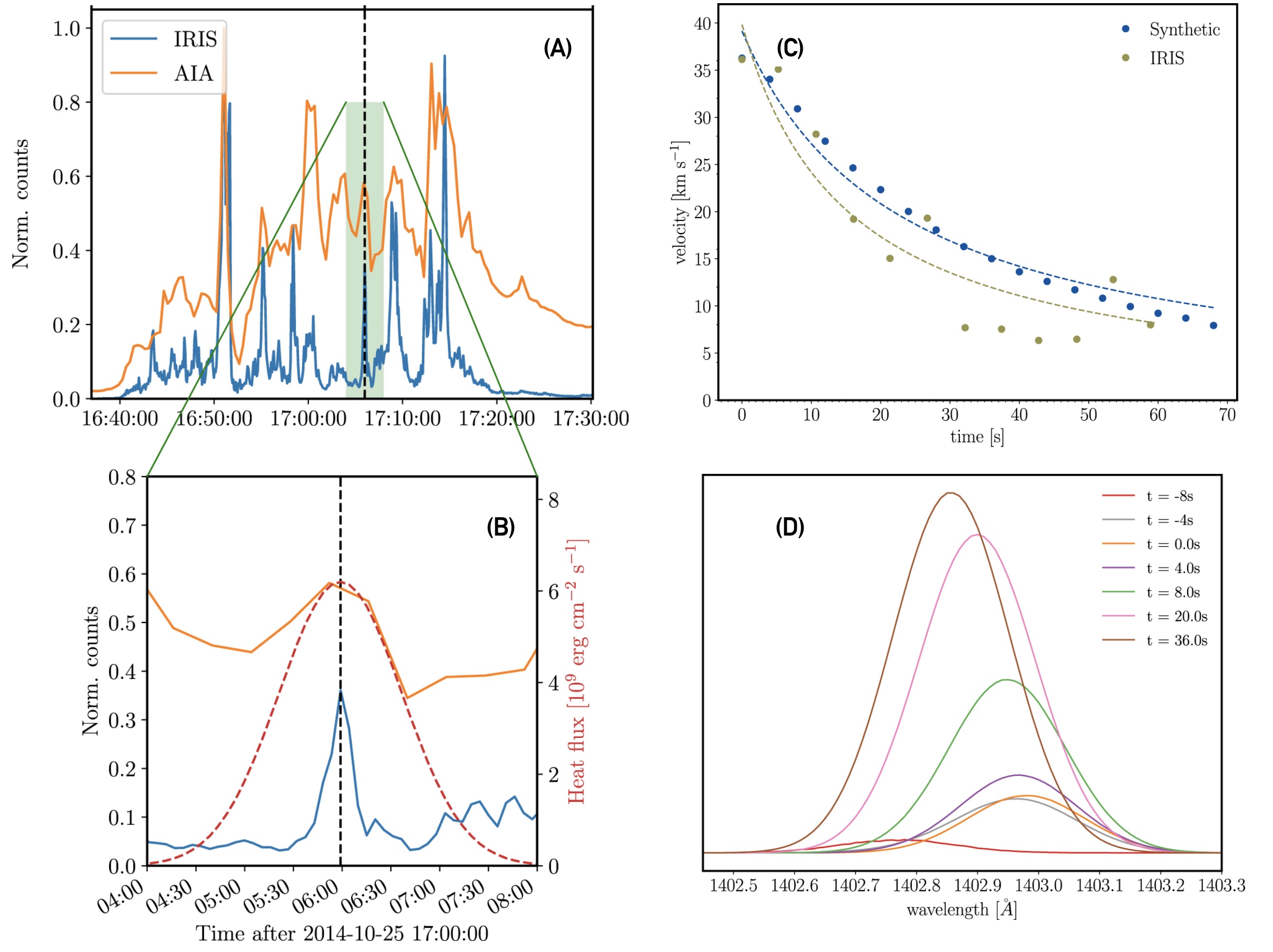}% This is a *.eps file
\end{center}
\caption{A thermal conduction driven flare simulation using \preft. Panel \textbf{(A)} shows observations of a single pixel in the 2014-Oct-25th X class flare. The \ion{Si}{iv} 1402~\AA\ line lightcurve is shown in blue, and the SDO/AIA 1600~\AA\ lightcurve is shown in orange. The dashed line shows the peak of the individual condensation event studied. Panel \textbf{(B)} shows a zoomed in view, where the injected heat flux is shown also, which was derived via the UFC method (red, dashed line). The resulting \ion{Si}{iv} redshifts due to the condensation are shown in panel \textbf{(C)} where blue symbols are the \preft\ results and green are the IRIS observations of that pixel. The dashed lines are fits to the decay of the condensation according to the model described in \cite{2021ApJ...912...25A}. Panel \textbf{(D)} shows the synthetic line profiles on an arbitrary intensity axis. Figure adapted from \cite{2022ApJ...926..164A}. \copyright AAS. Reproduced with permission.} \label{fig:preftfig}
\end{figure}

Applying their findings to an actual flare, \cite{2022ApJ...926..164A} analysed the 2014-October-25th X-class event. During the flare there were persistent redshifts of \ion{Si}{iv}, with $v_{Dopp} \sim 10$~km~s$^{-1}$, but on top of which were many transient ($<1$~minute) redshifts of several tens of km~s$^{-1}$. After carefully analysing the IRIS \ion{Si}{iv} spectra to extract a candidate condensation event to model, they used SDO/AIA and HMI data to trace a magnetic loop. Imaging the hard X-rays from RHESSI in the range 25-50~keV revealed only a coronal source, with no evidence of non-thermal particles at the footpoints. Thus, they used \preft\ to model this event as a `thermal' flare, to confirm that the observationally derived energy flux could drive the observed condensation. The density scale height was estimated from $u_{0}$ and $\tau_{1/2}$ as $H = 369$~km, which was used to initialise a rigid flux tube of length $85$~Mm. Since no evidence of electron precipitation into that footpoint was present, \cite{2022ApJ...926..164A} used the UV Footpoint Calorimeter method \citep[UFC;][]{2012ApJ...752..124Q,2013ApJ...770..111L,2018ApJ...856...27Z,2021ApJ...909...99Q} to determine the time-dependent energy flux injected into the loop. In that method, the energy flux is proportional to the intensity in the SDO/AIA 1600~\AA\ passband, with a scaling factor constrained by the SDO/AIA EUV channels and GOES Soft X-ray channels. The Enthalpy-Based Thermal Evolution of Loops \citep[EBTEL;][]{2008ApJ...682.1351K} model was used to synthesise the EUV and soft X-ray intensities for different energy fluxes, from which the best-fit match to the observations is used to define the scaling constant. EBTEL is a 0D model, and thus very quick to run, allowing many thousands of possible solutions to be generated efficiently to obtain the best-fit match. The 1600~\AA\ peak was fit by a Gaussian function, to obtain an energy flux profile with peak flux $F=6.2\times10^{9}$~erg~s$^{-1}$~cm$^{-2}$, and duration of a few minutes. The heating profile and observed lightcurves are shown in the lefthand column of Figure~\ref{fig:preftfig}.

Injecting this derived energy flux profile into the \preft\ loop resulted in the formation of a condensation, peaking prior to the peak of the energy input, and reaching a maximum of velocity $45$~km~s$^{-1
}$, before decreasing to $10$~km~s$^{-1}$, again prior to the peak energy input. Only a small fraction of the input energy flux was required to drive the condensation, leading to the conclusion that the timescales of condensations do not necessarily impart knowledge of the duration of energy input, in agreement with the analytical work of \cite{1989ApJ...346.1019F}. From the dynamic \preft\ simulation, the \ion{Si}{iv} spectral lines were synthesised assuming optically thin emission, and summed through the extent of each leg of the loop. Non-equilibrium effects were considered by tracking the change in ionisation state for each Lagrangian grid cell. While qualitatively consistent with the observed \ion{Si}{iv} emission from the footpoint, with similar redshifts produced (see the righthand column of Figure~\ref{fig:preftfig}), the synthetic intensities did not track the condensation evolution and were over an order of magnitude too high. The synthetic profiles were also fully redshifted rather than showing two component behaviour, possibly pointing to the need for many strands along the lines of \cite{2018ApJ...856..149R}'s multi-threaded modelling. Since the synthetic profiles continued to increase in intensity after their peak redshift, it is possible that the Gaussian form assumed for energy input was not correct. Keeping the total energy from the UFC method, but having a more impulsive initial energy release could decrease the energy deposited later in the event, reducing the line's intensity. Prior experiments with \preft\ that included the loop retraction exhibited much more rapid energy release \citep[][]{2015ApJ...813..131L}. Of particular importance here, aside from testing the work of \cite{2021ApJ...912...25A}, and demonstrating that the observed conductions could be produced in a thermal conduction-driven flare, was that this was the first simulation of a flare in which the energy input to the chromosphere was inferred from coronal observations.

%%%%%%%%%%%%%%%%%%%%%%%%%%%%%%%%%%%%%%%%%%%%%
%% FUTURE EFFORTS AND REMAINING CHALLENGES
%%%%%%%%%%%%%%%%%%%%%%%%%%%%%%%%%%%%%%%%%%%%%

\section{Future Directions}\label{sec:conc}

In this extensive review (including Paper 1) I have illustrated how the powerful combination of high spatial-, temporal-, and spectral- resolution observations of the chromosphere, transition region and corona, coupled with state-of-the-art numerical loop models can greatly further our understanding of the physics of solar flares. IRIS observations have been used to challenge the predictions of models, requiring us to update the physics we include in our models, as well as the ways in which we perform model-data comparisons. The models, on the other hand, have been used to assist in the interpretation of IRIS observations, particularly of the complex formation properties of the optically thick chromospheric and transition region spectral lines. Through study of Doppler shifts and line asymmetries, both their magnitude and lifetimes, we have come to understand the likely requirement of muti-threaded modelling to understand upflows. Chromospheric condensations, on the contrary, seemingly do not require multi-threaded modelling, presenting a discrepancy. Detailed comparisons of synthetic optically thick lines, and the NUV and optical continua, have demonstrated that while flare models can capture the upper chromospheric response to flare energy injection, we are perhaps missing heating deeper in the atmosphere. If higher energy fluxes are required, as suggested by some model-data comparisons, do those originate solely from non-thermal electrons or do we need other agents by which flare energy can be transported to the chromosphere? We must seriously consider Alfv\'enic waves, and non-thermal ions, and work to characterise the magnitude of energy directly deposited into the corona that is subsequently conducted down. Recent model improvements have included incorporating area expansion and suppression of thermal conduction, but more work on those features is required in particular to tie down the appropriate parameter spaces to use in general flare models (e.g. when producing grids of models). Modelling IRIS observables has also revealed that opacity effects could occur for transition region lines, which are important contributors to the `optically thin' radiative loss functions in the models. A re-evaluation of the loss tables would be a worthwhile enterprise considering their importance, with \hydrad\ NEI results helping to assess which species are also likely to suffer from non-equilibrium effects.  

There still remain open questions, likely pointing to missing ingredients in our models. While they are many, I take the liberty to note the questions that capture my own focus at the present moment: (1) What causes the very broad chromospheric spectral lines? Could this point to the need for deeper heating through the lower atmosphere, and if so what transports that energy? (2) What is the nature of white light flares? Are they largely of chromospheric origin like the Balmer continuum that IRIS has observations have suggested, or do we need additional heating through the lower atmosphere? (3) What is the source of post-impulsive phase energy transport that maintains the flare gradual phase? (4) What is the nature and magnitude of turbulence throughout the flaring atmosphere? Turbulence could explain broadening of coronal lines and the suppression of thermal conduction, as well causing heating in its own right and particle acceleration. Continued coupling of flare loop modelling and high quality observations, can help address, and close, these questions. A powerful method to both guide and evaluate the models is the atmospheric stratification obtained from spectral inversions, for example using the IRIS2 database of \ion{Mg}{ii} inversions \citep[][]{2019ApJ...875L..18S} and subsequent updates to further constrain the atmospheres using other IRIS lines. These should be compared to the output of flare models, with the results of both critically compared and interrogated to determine when they agree, when they disagree and in that event, crucially, why they disagree. 

IRIS continues to deliver excellent observations. Both from observational analysis, and from numerical modelling it has become clear that flare processes can occur at rapid cadence, on the order of seconds to sub-second. In the present solar cycle there will be concerted efforts to obtain flare observations of the main IRIS lines (\ion{Mg}{ii} k, \ion{Si}{iv} 1402~\AA, \ion{C}{ii} resonance lines, and sometimes \ion{O}{i} 1356~\AA) with very high cadences of $t<1.5$~s or even at sub-second cadence. This observing campaign has already caught many flares\footnote{\url{https://iris.lmsal.com/data.html}}. Additionally, we can look ahead to future missions that build upon the heritage of IRIS, both technologically and from the methodological approach of strongly coupling modelling and observations. Those missions include the Solar-C EUV High-Throughput Spectroscopic Telescope \citep[EUVST;][]{2019SPIE11118E..07S}, and the Multi-slit Solar Explorer \citep[MUSE;][]{2020ApJ...888....3D}. EUVST is a single-slit spectrometer with a huge temperature coverage spanning the chromosphere through hot flare plasma ($T = 0.02-15$~MK), observing with spatial resolution of $0.4$~arcsec and temporal resolution down to $\sim2$~s in sit-and-stare mode (of course lower cadence longer if rastering). MUSE is an also an EUV observatory, but is specifically designed with temporal cadence in mind, such that it has 37 slits enabling rastering over an active region field of view in only 12s with $0.4$~arcsec spatial resolution, and $1$~s cadence in sit-and-stare mode. Fewer lines are observed by MUSE given the complexity of its 37 slit design, but those that are observed sample the transition region and corona. With 37 slits we will no longer have the frustration of missing the interesting features because the slit was pointed in the `wrong' place. We will instead have transition region and coronal imaging spectroscopy over much larger areas than ever achieved. Recent modelling efforts demonstrated the transformative science that MUSE will achieve in the area of solar flares and eruptions \citep{2022ApJ...926...53C}. Coordination of these space based missions with ground-based observatories such as Big Bear Solar Observatory (BBSO), the Swedish Solar Telescope (SST), GREGOR, and the Daniel K. Inouye Solar Telescope (DKIST) should provide coverage from photosphere through corona, to observe the full flaring atmosphere.

At the same time as obtaining ever higher quality observations we must strive to improve our modelling ability and frameworks. To make rapid and serious progress not only in solar flare physics, but in modern space physics more generally, we must have significant investment of time and resources in models. This involves continued improvements of our existing field-aligned models \citep[some examples are discussed in section 5.2 of][]{2022ApJ...928..190K}, but also to attempt to bridge the gap to multi-dimensional models. A fully 3D radiative-MHD model capable of simulating an NLTE chromosphere to sub-metre scales following flare energy injection is a daunting task that may not be fully realised for many years, but certainly we must attempt to include multi-dimensional effects. Efforts are already underway to study 2D/3D RT effects. For example, \cite{2022MNRAS.516.6066O} are exploring the effects of 2D radiation transfer, using 1D field-aligned \radyn\ models embedded within regions of quiet-Sun, finding significant differences may be present (including on the intensity of transition region lines, \textsl{private communication}). Additionally, non-thermal electrons have now been included in 3D radiative-MHD simulations, with very low fluxes to explore nano-flare heating \citep[][]{2018A&A...620L...5B}, and hopefully larger fluxes typical of flares may be explored in the future. However, we must also think about how to handle horizontal expansion of plasma in flare footpoints in the field-aligned models (that can handle the small vertical scales required in the chromosphere during flares). 

Recently, Dr. Joel Allred and co-authors proposed an end-to-end modelling framework of solar eruptive events in a white paper titled `Next-Generation Comprehensive Data-Driven Models of Solar Eruptive Events' \footnote{White papers will appear in the Bulletin of the American Astronomical Society, but are for now hosted by the NAS on their website: \url{https://www.nationalacademies.org/our-work/decadal-survey-for-solar-and-space-physics-heliophysics-2024-2033}. A link to Dr. Allred's white paper is: \url{http://surveygizmoresponseuploads.s3.amazonaws.com/fileuploads/623127/6920789/140-a8d175a52b5836b620abf6d961febf97_AllredJoelC.pdf}} submitted to the National Academy of Science Solar and Space Physics Decadal Survey. In this framework, models that target  different aspects of solar eruptive events should be linked such that the output of one is either the input to the next link in the chain, or strongly guides/constrains the next link, ideally in a data-driven or data-constrained manner. Field-aligned models would be an essential component. An example of such a chain could be: (1) an MHD model of the build up and release of magnetic energy, that then drives (2) a model of particle acceleration, the energy spectrum of which drives, (3) a model that propagates and dissipates non-thermal particles, producing (4) the radiative hydrodynamic response of the atmosphere, that (5), ultimately predicts observables, including geometry of the original magnetic field/loops (e.g. via a \radynarcade\ type framework). The latter three steps are of course already being done with the models discussed in this review (e.g. \radyn+\fp). Joined up modelling such as that described by Allred et al, interrogated by high-resolution observations, is perhaps the best way to make rapid progress whilst covering the vast range of scales, from MHD to kinetic, present in solar flares. 

%%%%%%%%%%%%%%%%%%%%%%%%%%%%%%%%%%%%%%%%%%%%%
%% ENDMATTER
%%%%%%%%%%%%%%%%%%%%%%%%%%%%%%%%%%%%%%%%%%%%%

\section*{Conflict of Interest Statement}
The author declares that the research was conducted in the absence of any commercial or financial relationships that could be construed as a potential conflict of interest.

\section*{Author Contributions}

GSK performed the literature review and wrote the manuscript. 

\section*{Funding}
GSK acknowledges funding via a NASA ROSES Early Career Investigator Award (Grant\# 80NSSC21K0460), and the Heliophysics Supporting Research program (Grant\# 80NSSC21K0010). 

\section*{Acknowledgments}
IRIS is a NASA small explorer mission developed and operated by LMSAL with mission operations executed at NASA Ames Research center and major contributions to downlink communications funded by the Norwegian Space Center (NSC, Norway) through an ESA PRODEX contract. This manuscript benefited from discussions held at a meeting of International Space Science Institute team: “Interrogating Field-Aligned Solar Flare Models: Comparing, Contrasting and Improving,” led by Dr. G. S. Kerr and Dr. V. Polito. I also thank the following colleagues for their help, and patience, with answering questions related to \radyn, \preft, \flarix, and \hydrad: Dr. Joel Allred, Dr. Mats Carlsson, Dr. Adam Kowalski, Dr. Vanessa Polito, Dr. Dana Longcope, Dr. William Ashfield, Dr. John Unverferth, Dr. Stephen Bradshaw, Dr. Jeffrey Reep, Dr. Jana {Ka{\v{s}}parov{\'a}, Dr. Petr Heinzel, and Dr. Michal Varady. Finally, I thank the referees for their diligent reading and useful comments which improved both content and clarity of this review.

\bibliographystyle{frontiersinSCNS_ENG_HUMS} 

\bibliography{Kerr_Frontiers_IRISFlareReview_LoopModels}

\end{document}